\documentclass[12pt,preprint]{aastex}
\newcommand{\msun}{{M}_{\odot}}
\newcommand{\Hm}{\rm{H}^{-}}
\newcommand{\me}{\rm{e^{-}}}
\newcommand{\Hp}{\rm{H}^{+}}
\newcommand{\mH}{\rm{H}}
\newcommand{\mHt}{\rm{H}_{2}}
\newcommand{\mHtp}{\rm{H}_{2}^{+}}
\newcommand{\hii}{\hbox{H\,{\sc ii}}\,}
\newcommand{\mC}{\rm{C}}
\newcommand{\cp}{\rm{C^{+}}}
\newcommand{\mO}{\rm{O}}
\newcommand{\op}{\rm{O}^{+}}
\newcommand{\mSi}{\rm{Si}}
\newcommand{\sip}{\rm{Si^{+}}}

\def\simless{\mathbin{\lower 3pt\hbox
   {$\rlap{\raise 5pt\hbox{$\char'074$}}\mathchar"7218$}}}   
\def\simgreat{\mathbin{\lower 3pt\hbox  
   {$\rlap{\raise 5pt\hbox{$\char'076$}}\mathchar"7218$}}} 

\bibpunct{(}{)}{;}{a}{}{,}

\shorttitle{
Cooling of Ionized Gas in Small Protogalactic Halos}
\shortauthors{Jappsen et al.}

\begin{document}

\title{Star Formation at Very Low Metallicity. II: On the 
Insignificance of Metal-Line Cooling During the Early Stages 
of Gravitational Collapse}

\author{A.-K. Jappsen\altaffilmark{1,2}, S.~C.~O. Glover\altaffilmark{2,3}, R.~S. Klessen\altaffilmark{4,2}, 
M.-M. Mac Low\altaffilmark{3}}

\altaffiltext{1}{Canadian Institute for Theoretical Astrophysics,\\
University of Toronto, Toronto, ON M5S 3H8, Canada;
jappsen@cita.utoronto.ca}
\altaffiltext{2}{Astrophysikalisches Institut Potsdam,\\14482
 Potsdam, Germany; sglover@aip.de}
\altaffiltext{3}{Department of Astrophysics, American Museum of Natural History,
\\New York, NY 10024-5192; 
mordecai@amnh.org}
\altaffiltext{4}{Zentrum f\"ur Astronomie der Universit\"at Heidelberg, 
Institut f\"ur Theoretische Astrophysik, \\69120 Heidelberg, Germany;
rklessen@ita.uni-heidelberg.de}

\begin{abstract}
We study the influence of low levels of metal enrichment on the cooling 
and collapse of ionized gas in small protogalactic halos using 
three-dimensional, smoothed particle hydrodynamics simulations. 
Our initial conditions represent protogalaxies forming within a fossil 
$\hii$ region, a previously ionized $\hii$ region which has not yet had 
time to cool and recombine. Prior to cosmological reionization, such regions 
should be relatively common, since the characteristic lifetimes of the likely 
ionizing sources are significantly shorter than a Hubble time. We show that in 
these regions, $\mHt$ is the dominant and most effective coolant, and 
that it is the amount of $\mHt$ formed that determines whether or not
the gas can collapse and form stars. 

At the low metallicities ($Z < 10^{-3} \: {Z_{\odot}}$)
thought to be associated with the transition from Population III
to early Population II star formation, metal-line cooling has an
almost negligible effect on the evolution of low-density gas, 
altering the density and temperature evolution of the gas by 
less than 1\% compared to the metal-free case at densities below 
$1\,\mathrm{cm^{-3}}$ and temperatures above $2000\,\mathrm{K}$. 
Although there is evidence that metal-line cooling becomes more
effective at higher density, we find no significant differences
in behaviour from the metal-free case at any density below our
sink particle creation threshold at $n = 500 \: {\rm cm^{-3}}$.
Increasing the metallicity also increases the importance of 
metal-line cooling, but it does 
not significantly affect the dynamical evolution of the low-density gas until ${Z} \sim 0.1 \: {Z_{\odot}}$.
This result holds regardless of whether or not an ultraviolet
background is present.

\end{abstract}
\section{Introduction}
Within the framework of cold dark matter (CDM) cosmology the formation of
structure proceeds in a hierarchical fashion. At high redshifts, low-mass
halos with virial temperatures less than $\sim 10^4\,\mathrm{K}$
are abundant. The cooling of primordial gas in these halos is regulated by 
molecular hydrogen, as $\mHt$ is the only coolant present
in significant quantities that remains effective at temperatures below
$10^4\,\mathrm{K}$ \citep{SAS67, PEE68, MAT69}. \citet{TEG97} developed
analytic methods to model early baryonic collapse via $\mHt$ cooling. Numerical 
studies of the formation of
primordial gas clouds and the first stars indicate that this process likely
began as early as $z \sim 30$ \citep{ABE02,BRO02}. \citet{YOS03} further
utilized simulations to develop a semi-analytic model based on the Tegmark et
al. (1997) methods and included the effects of dynamical heating caused by the
thermalization of kinetic energy of infall into a deepening potential. Both
approaches suggest that only gas in halos more massive than some critical mass
$M_{\mathrm{crit}}$ will cool effectively. Much of this work has recently been
reviewed by \citet{BRO04}, \citet{CIA05}, and \citet{GLO05}.

Population {\sc III} stars are the first potential producers of UV photons that can
contribute to the reionization process and are the first producers of the
metals required for the formation of population {\sc II} stars.  
An important question is whether small protogalaxies that formed within the relic
\hii regions left by these first stars could form stars themselves, or whether the
elevated temperatures and fractional ionizations found in these regions suppressed
star formation until larger protogalaxies formed. In a recent analytical study, \citet{OH03}
argue that the first stars injected sufficient entropy into the early
intergalactic medium (IGM) by photoionization heating and supernova explosions to
strongly suppress further star formation inside low-mass protogalactic halos in the affected
regions. On the other hand, previous numerical work by \citet{RIC02} and \citet{OSH05}, 
as well as our own simulations of hot, ionized gas in small protogalactic halos, show that 
gravitational collapse mediated by $\mHt$ cooling remains possible in such regions, within 
protogalaxies with masses above a certain threshold, implying that in these systems star 
formation is not strongly suppressed.

A further form of  negative feedback that must be taken into account comes from UV photons 
within the Lyman-Werner bands of $\mHt$, which are produced in large numbers by massive 
population {\sc III} stars and which can photodissociate $\mHt$, thereby quenching molecular 
hydrogen cooling and delaying the cooling and collapse of the primordial gas 
\citep{HAI00, MAC01, GLO01,JOH06}.  

Metals produced by the first stars will also be injected into some fraction of the ionized volume,
and so will be present in gas falling into new or existing protogalactic halos.
The question then arises as to how this low level of metal enrichment affects the ability of the 
gas to cool and collapse. \citet{BRO01} studied the collapse of cold, metal-enriched gas and
argued that there exists a critical metallicity $Z_{\mathrm{crit}}$ below which the 
gas fails to undergo continued collapse and fragmentation. However, 
their simulations did not allow for $\mHt$ formation and $\mHt$ cooling, which is 
expected to play an important role, unless the external UV background is extremely 
strong.  Also, by starting with cold gas, they implicitly assume that no extra entropy or energy
has been added to the gas during its enrichment; although, as \citet{OH03} have 
shown, this is unlikely to be the case. We reexamine this
question using more appropriate initial conditions and a more detailed treatment
of the cooling and chemistry of the metal-poor gas.

We present simulations showing that metals play a dominant role in regulating
the cooling of ionized gas in small protogalactic halos only for metallicities
${Z} \sim 0.1 \: {Z_{\odot}}$ or greater. At lower metallicities, the
metals have little effect and the question of whether or not gas in a 
particular halo collapses is entirely determined by the amount of $\mHt$ 
forming in that halo. We reach the same conclusion regardless of whether or 
not the gas is illuminated by a Lyman-Werner UV background, as in the absence 
of $\mHt$ the metals themselves do not provide enough cooling to allow the 
gas to collapse.

\section{Simulations}
\subsection{Numerical Method}
To help us to assess the influence of metals on the cooling and collapse of
gas in small protogalactic halos, we performed a number of numerical 
simulations. During collapse, gas increases in density by several orders of
magnitude, and so is best simulated by a numerical method with a high
dynamical range. We therefore chose smoothed particle hydrodynamics
(SPH). Excellent overviews of the method, its numerical implementation, and
some of its applications are given in reviews by \citet{BEN90} and
Monaghan(1992, 2005). We use version 1 of the parallel code GADGET, 
designed by \citet*{SPR01}.  SPH is a Lagrangian method for simulating 
astrophysical flows, in which the fluid is represented by an ensemble 
of particles, with flow quantities at a particular point obtained by 
averaging over an appropriate subset of neighboring SPH particles.
\nocite{MON92, MON05}
As particle time steps in an SPH code are constrained by the Courant condition,
they grow increasingly short as more particles cluster in high-density regions.
Replacing dense cores with artificial sink particles can therefore lead to a
considerable increase in computational performance, allowing the dynamical
evolution of the gas to be followed over many free-fall times. In the runs presented 
here we introduce sink particles according to the prescription of \citet{BAT95}.
Sink particles are created once the density rises above $500\,\mathrm{cm^{-3}}$ 
and are endowed with an accretion radius of $5\,\mathrm{pc}$. On every time step, 
any gas within this accretion radius that is gravitationally bound to the sink particle 
is accreted by it.  The design and implementation of our sink particle algorithm is 
discussed in more detail in \citet{JAP05}.  

\subsection{Chemistry and Cooling}
\label{chemcool}
We have further modified GADGET to allow us to follow the nonequilibrium
chemistry of the major coolants in both primordial and low-metallicity gas.
Our chemical model is a simplified version of the general model for 
low-density, low-metallicity gas presented in Paper I \citep{gj07}. In our
simplified network, we follow the chemistry of twelve separate species: 
free electrons, $\Hp$, $\Hm$, $\mH$, $\mHtp$, $\mHt$, $\mC$, ${\rm C^{+}}$, 
$\mO$, ${\rm O^{+}}$,  $\mSi$, and ${\rm Si^{+}}$.  The more detailed 
model in Paper I also follows the chemistry of ${\rm He}$, ${\rm He^{+}}$,
${\rm Si^{++}}$, D, ${\rm D^{+}}$, and HD. However, the helium and 
${\rm Si^{++}}$ chemistry are important primarily at temperatures 
$T \simgreat 10^{4} \: {\rm K}$, or if the gas is partially ionized by 
X-rays or cosmic rays, while the deuterium chemistry is only significant
at temperatures $T < 200 \: {\rm K}$, which are only reached in a few of
our simulations, and so the neglect of these additional reactions 
reduces the computational cost imposed by the chemistry without 
significantly affecting our main results. The other simplification we 
made here compared to the model of Paper I is the neglect of 
photochemical reactions involving photons more energetic than 
$13.6 \: {\rm eV}$; specifically, we omit the photoionization of $\mH$,
He, $\mHt$, O, and ${\rm Si^{+}}$ (reactions 48, 50, 54, 57, and 59 in Paper I).  
We omit these reactions because we assume that during the simulations,
our model protogalaxies are shielded by the neutral hydrogen in the 
intergalactic medium from any external sources of ionizing radiation. 
A full list of reactions included in the chemical network used in the
simulations presented in this paper is given in Table~\ref{tab:chem_gas}. 
Further details (including a discussion of why these particular reactions 
were chosen and full details of the rate coefficients used) are given in Paper I.

In implementing this chemical network within GADGET, we made the
decision to follow the nonequilibrium abundances of $\mHt$, $\Hp$, 
${\rm C^{+}}$, ${\rm O^{+}}$, and ${\rm Si^{+}}$ and use conservation 
laws for charge and element abundance to track  ${\rm H}$, $\me$, $\mC$, $\mO$, 
and $\mSi$. The remaining two species, the $\Hm$ and $\mHtp$ ions, 
have very short chemical equilibrium timescales and hence were 
assumed, for simplicity, to reach equilibrium instantaneously.
To compute the rate of photochemical reactions (such as the photodissociation
of $\mHt$) in runs in which an ultraviolet background is present, we assume a 
background spectrum that has the shape of a $10^{5} \: {\rm K}$ black body 
(as should be typical of the brightest Population III stars; see e.g.\ 
\citealt{coj00}), cut off at energies greater than $13.6 \: {\rm eV}$ to
account for absorption by neutral hydrogen in the IGM. We 
quantify the strength of the background by fixing the flux at the Lyman limit: 
$J(\nu_{\alpha}) = 10^{-21} J_{21} \: {\rm erg} \: {\rm s^{-1}} \: {\rm cm^{-2}} 
\: {\rm Hz^{-1}} \: {\rm sr^{-1}}$. A list of the resulting photochemical rates 
is given in Table~\ref{tab:chem_gas_photo}, computed using the 
cross sections listed in Paper I.

If sufficient $\mHt$ forms within the protogalaxy, it will begin to
self-shield, reducing the effective photodissociation rate. An exact
treatment of the effects of self-shielding is computationally
infeasible, as it would require us to solve for the full spatial,
angular, and frequency dependence of the radiation field at every
time step. Instead, we have chosen to incorporate it in an approximate
manner. We assume that the dominant contribution to the self-shielding
at a given point in the protogalaxy comes from gas close to that
point, and so we only include the contribution to the self-shielding that
comes from nearby $\mHt$. 

Finally, we assume that ionization from X-rays or cosmic rays is
negligible. Previous work suggests that even if a low level
of ionization from such sources is present, it will not have a major effect on the
outcome of the collapse \citep{GB03, MAC03}. In any case, the initial fractional
ionization in our simulations is very much higher than could be produced by a realistic
flux of X-rays or cosmic rays. 

\begin{figure}
\centering
\includegraphics[width=20pc,angle=270,clip=]{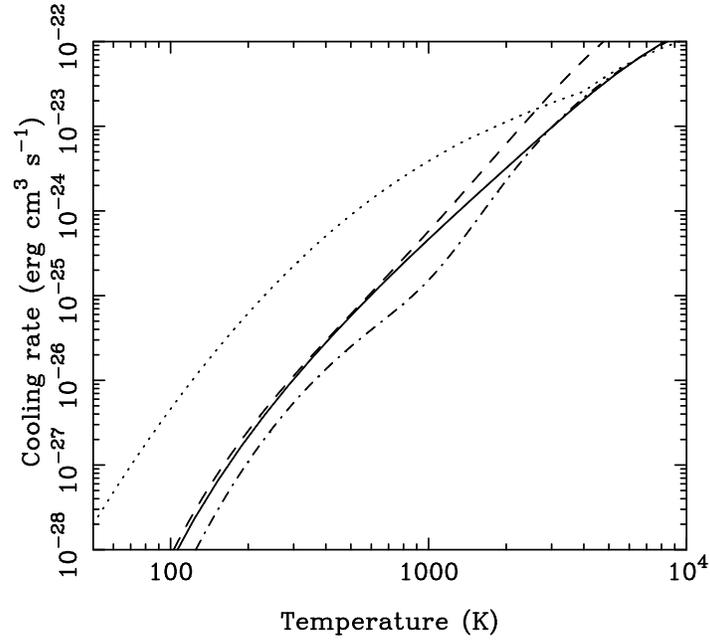}
\caption{Comparison of various parameterizations of the $\mHt$ cooling 
function, plotted in units of ${\rm erg} \: {\rm cm}^{3} \: {\rm s}^{-1}$.
Rates are computed for a number density $n = 1 \: {\rm cm^{-3}}$ and assuming
that $n_{\mH} \gg n_{\mHt}$ and that the $\mHt$ ortho-to-para ratio is 3:1.
{\it Solid line}: Le~Bourlot et al.~(1999); 
{\it dashed line}: Galli \& Palla~(1998); {\it dotted line}: Lepp \& Shull~(1983); 
{\it dash-dotted line}: Hollenbach \& McKee~(1979).}
\label{h2cool}
\end{figure}

A second major modification that we have made to the GADGET code is
a treatment of radiative heating and cooling. In primordial gas, we include 
cooling from three main sources: electron impact excitation
of atomic hydrogen (Ly$\alpha$ cooling), which is
effective only above about $8000 \: {\rm K}$; rotational and
vibrational excitation of $\mHt$; and Compton
cooling. Rates for Ly$\alpha$ cooling and Compton cooling were
taken from \citet{CEN92}, while for $\mHt$ rovibrational cooling we
used a cooling function from \citet{BOU99}. A number of different 
parameterizations of the $\mHt$ cooling function have been used in
the astrophysical literature, and so we show for reference in 
Figure~\ref{h2cool} how the \citet{BOU99} rate compares with
various of the other rates.

In addition, in cases where the gas is metal enriched, we compute the amount of 
fine structure cooling coming from $\mC$, ${\rm C^{+}}$, $\mO$, $\mSi$, and ${\rm Si^{+}}$.
In this calculation, we include the effect of excitation of the fine-structure lines by
the CMB, which prevents the gas from cooling below $T_{\rm CMB}$. We also include
heating from $\mHt$ photodissociation, following \citet{BLA77}, and from
the ultraviolet pumping of $\mHt$, following \citet{BUR90}. 

We do not include cooling from HD, as this is only significant compared to
$\mHt$ at temperatures below $T \sim 200 \:{\rm K}$. Dense gas which reaches 
these temperatures in our simulations is very soon thereafter accreted by a sink 
particle, at which point its further temperature evolution cannot be followed.
We therefore do not expect that the lack of HD cooling in our models will 
significantly affect our conclusions. We also omit cooling from other primordial 
molecules and molecular ions (e.g.\ ${\rm LiH}$ and ${\rm H_{3}^{+}}$), as these
are unimportant at the densities and temperatures encountered in our
simulations. Heavy molecules such as ${\rm CO}$ are also omitted, as they 
form in significant abundances only at densities greater than those considered 
in this paper, as has already been discussed at length in Paper I.
A brief list of the thermal processes included in our simulations is given in 
Table~\ref{cool_model}; further details, in particular of our treatment of fine-structure cooling, can be found in Paper I. 

\subsection{Initial Conditions}
\label{IC}
\subsubsection{Initial temperature and density distribution}
We initialize each of our simulations with gas that is hot 
($T_{\mathrm{g}}=10^4\:$K) and fully 
ionized. The physical situation that these initial conditions are intended to 
represent is a protogalaxy forming within what \citet{OH03} term a \"fossil\"
\hii region, an \hii region surrounding an ionizing source that has 
switched off, containing gas that has not yet had time to cool and 
recombine. Prior to cosmological reionization, such regions should be 
relatively common, since the characteristic lifetime of the likely ionizing 
sources -- massive Population {\sc III} stars and/or active galactic nuclei 
-- are significantly shorter than the Hubble time.

The initial temperature of gas in such a fossil \hii region should,
strictly speaking, depend on various factors such as the spectrum
and strength of the ionizing source and the density of the gas.
However, since the gas cools rapidly through Ly$\alpha$ cooling
at the start of our simulations, and since the temperature at which 
Ly$\alpha$ cooling becomes ineffective does not depend on the 
initial temperature of the gas, we do not expect our results to be sensitive
to small changes in the initial value of $T_{g}$.

For our purposes, we do not need to follow  the assembly 
history of the dark matter halo in which the protogalaxy resides, or to
study the response of the dark matter halo to the cooling of the gas. Therefore, 
we choose to model the influence of the dark matter halo (hereafter simply the `halo') by
using a fixed background potential. To construct this potential, we assume
that the halo is spherically symmetric, with the density profile of \citet*{NAV97}:
\begin{equation}
\rho_{\mathrm{dm}}(r)=\frac{\delta_{\mathrm{c}}\rho_{\mathrm{crit}}}{r/r_{\mathrm{s}}(1+r/r_\mathrm{s})^2},
\label{nfw-profile}
\end{equation}
where $r_{{s}}$ is a scale radius, $\delta_{{c}}$ is a
characteristic (dimensionless) density, and $\rho_{\mathrm{crit}}=3H^2/8\pi G$
is the critical density for closure. Note that we use a value for the Hubble 
constant of $H_0 = 72\,\mathrm{km}\,\mathrm{s}^{-1}\,\mathrm{Mpc}^{-1}$ \citep{SPE03}.
Following \citet{NAV97}, we calculate the characteristic density and the 
scale radius from a given redshift and dark halo mass. We truncate the halo 
at the radius at which the value of $\rho_{\mathrm{dm}}$ given by 
equation~\ref{nfw-profile} equals the cosmological background density at 
the beginning of the simulation. To simplify the discussion of our simulation
results, we take as a fiducial example a halo with a total mass 
$M_{\rm tot} = 7.8 \times 10^{5} \: {M_{\odot}}$ and an initial redshift 
$z_{i} = 25$. For this example halo, the virial temperature 
$T_{\rm vir} = 1900 \, \mathrm{K}$, the virial radius $r_{\rm vir} = 0.1 \, {\rm kpc}$,
and the truncation radius $r_{{t}} = 0.49\:\mathrm{kpc}$, where both
radii are given in physical units. The scale radius $r_{{s}}$ of this 
example halo is $29\,\mathrm{pc}$, and the full computational volume is a box 
of side length $1\,\mathrm{kpc}$. In addition to this fiducial example, we
also studied cooling and collapse in halos with various different masses
in the range $4.0 \times 10^{4} \: M_{\odot} < M_{\rm tot} < 1.2 \times 10^{7} \: M_{\odot}$, 
at redshifts in the range $20 < z < 30$. A complete list of runs is given in 
Table~\ref{tab:run_params}.

We use periodic boundary conditions for the
hydrodynamic part of the force calculations to keep the gas bound within the
computational volume. The self-gravity of the gas and
the gravitational force exerted by the dark matter potential are not
calculated periodically, since we assume that other dark matter halos and their
gas content are distant enough to neglect their gravitational influence. 

As we are dealing with a gravitationally bound system (the dark matter halo)
that has decoupled from the Hubble flow, we do not include the effects of
cosmological expansion, and we perform our simulations using physical rather
than co-moving coordinates. This simplification should be reasonably accurate
for gas within the virial radius of the halo, although it will break down
at larger radii. However, the behaviour of the gas at radii $r \gg r_{\rm vir}$
will have little effect on the evolution of the gas in the central regions 
of the halo, and since it is the latter that we are primarily 
concerned with in this paper, this simplification should be adequate for
our current purposes.

We begin our simulations with a uniform distribution of gas with an initial
density $\rho_{{g}}$, taken to be equal to the cosmological background
density, and then allow the gas to relax isothermally until it reaches hydrostatic
equilibrium. Note that this initial phase of the simulation is merely a convenient way
of generating the appropriate initial conditions for the simulation proper, and so 
we do not include the effects of chemical evolution or cooling during this phase.
The mass of gas present in our simulation was taken to be a fraction
$\Omega_{{b}}/\Omega_{\mathrm{dm}}$ of the total mass of dark matter, where the dark matter 
density $\Omega_{\mathrm{dm}}=\Omega_{{m}}-\Omega_{{b}}$, and where 
$\Omega_{{b}}$ is the baryon density and $\Omega_{{m}}$ is the matter density. We 
take values for the cosmological parameters from \citet{SPE03} of $\Omega_{{b}}=0.047$
and $\Omega_{\mathrm{m}}=0.29$, giving us a total gas mass of
$M_{{g}}=0.19\,M_{\mathrm{dm}}$. Here, $M_{\mathrm{dm}}$ is the sum of the halo mass and the 
mass of the dark matter background in the simulated volume. For our fiducial example, and for our example, $M_{\mathrm{dm}} = 1.84 \times 10^{6}\:{M_{\odot}}$. Therefore,
in this case $M_{{g}}=3.5 \times 10^{5}\:{M_{\odot}}$. The number of 
SPH particles used to represent the gas depends on the mass of the halo. We
selected an appropriate particle number and particle mass by requiring that 
the number of particles used, $N$, and the particle mass, $M_{\rm part}$,
satisfied $N \geq 10^{5}$ and $M_{\rm part} \leq 5.0 \: {M}_{\odot}$ respectively. 
Our SPH smoothing kernel encompasses approximately 40 particles and since we need 
twice this number in order to properly resolve gravitationally bound objects 
\citep{BAT97}, our mass resolution, in the worst case, is  $M_{\mathrm{res}} 
\simeq 80\,m_{\mathrm{part}} = 400\: {M_{\odot}}$. In some of the smaller 
halos we simulate, our mass resolution is significantly better. The values of
$N$ and $M_{\rm part}$ used for each of the different halo masses are summarized
in Table~\ref{tab:run_derived_vals}, along with the corresponding values of
$M_{\rm res}$ and also the  virial temperatures of the various halos.

At our sink creation density, $M_{\mathrm{J}}$ exceeds $M_{\mathrm{res}}$ as long as 
$T > 30 \: \rm{K}$. This temperature is much smaller than that reached by the gas 
in our simulations, and so we always satisfy the \citet{BAT97} resolution criterion 
by a comfortable margin in gas which has not been accreted by a sink.

\subsubsection{Metallicity}
We studied the evolution both of zero-metallicity, primordial gas and of 
metal-enriched gas. The range of metallicities studied was 
$10^{-3} < {Z} / {Z_{\odot}} < 1.0$. The value used in any particular
simulation is summarized in Table~\ref{tab:run_params}. In particular, we
performed a large number of simulations with ${Z} = 10^{-3}\: {Z}_{\odot}$.
This value is an upper limit derived from QSO absorption-line studies of the 
low column density Ly$\alpha$ forest at  $z \sim 3$ \citep{PET99}. 
Estimates of the globally-averaged metallicity produced by the sources 
responsible for reionization are also typically of the order of 
$10^{-3}\, {Z}_{\odot}$ \citep[see e.g.][]{RIC04}. Moreover, the fact that
several hundred stars with $[{\rm Fe}/{\rm H}] < -3$ have been observed in
the Galactic halo provides solid evidence that solar mass stars can form
at this metallicity \citep{chr03}. We originally intended to examine the behaviour
of the gas at even lower metallicities, but these additional simulations were
rendered unnecessary by the fact that even at ${Z} = 10^{-3} 
\: {Z_{\odot}}$ metal cooling proved to be of no importance at the low
gas densities studied in this paper.

In our simulations of metal-enriched gas, we assume that mixing is efficient and that 
the metals are spread out uniformly throughout the computational domain.
We also assume that the relative abundances of the 
various metals in the enriched gas are the same as in solar metallicity gas; given 
the wide scatter that exists in observational determinations and theoretical 
predictions of abundance ratios in very low-metallicity gas, 
this seems to us to be the most conservative assumption. However, variations in the 
relative abundances of an order of magnitude or less will not significantly alter our 
results. 

As we discuss in Paper I, substantial uncertainties exist concerning
the presence, abundance, size distribution, and composition of dust grains
in low-metallicity gas at high redshift. An exhaustive examination of the various
possibilities would require an unfeasibly large number of simulations to
be performed, and so we instead make the simplifying assumption that if dust
is present, it has similar properties to Milky Way dust, but with an abundance
that is a factor of ${Z}/{Z_{\odot}}$ smaller than the Galactic value. To
allow us to investigate the relative importance of the dust and the gas-phase
metals, we perform simulations both with and without dust.

\subsubsection{Other details}

We also investigate the influence of a UV background. By $z=25$, a
considerable UV background may already have developed \citep{HAI00,GB03},
particularly if cosmological reionization occurs at a redshift near the upper 
end of the range suggested by the {\it WMAP} polarization results 
\citep{sper06}. To explore how the presence of a UV background will influence our 
conclusions, we have run simulations with UV field strengths of $J_{21} = 0.01$ 
and 0.1, as well as with no background present.

Simulations were run for $\sim 1$ Hubble time, $t_{\rm H}$, which at $z =25$ 
corresponds to about $100 \: {\rm Myr}$. At times $t > t_{\rm H}$, individual
dark matter halos can no longer be treated as isolated objects, since most
halos will have undergone at least one major merger or interaction by this 
time. For our fiducial halo, the free-fall time $t_{\rm ff} \sim (G\rho)^{-1/2}$
varies between $4\,\mathrm{Myr}$ near the central cusp and $450\,\mathrm{Myr}$ in 
the outer envelope.

For ease of reference, we have given the runs discussed in the following sections
simple alphanumeric identifiers. We have (arbitrarily) divided up our model halos
into three size classes: ``small'' halos, with masses $M < 10^{5} \: \msun$; 
``intermediate'' halos, with $10^{5} < M < 4.5 \times 10^{6} \: \msun$; and 
``large'' halos with $M > 4.5 \times 10^{6} \: \msun$. The first character
of the identifier indicates with which of these classes the run in question is 
associated: ``S'' denotes a small halo, ``I'' an intermediate halo, and ``L'' 
a large halo. The next two digits in the identifier denote the initial redshift 
of the simulation. Next, there follows a character denoting the metallicity:
``Z''  corresponds to zero metallicity, while ``L,'' ``M,'' ``N,'' and ``S'' 
correspond to metallicities $10^{-3}$, $10^{-2}$, $10^{-1}$, and 1 ${Z_{\odot}}$, 
respectively. Following this is a number denoting the strength of the UV
background: 1 corresponds to no background, while 2 and 3 correspond to 
$J_{21} = 0.01$ and 0.1, respectively. Finally, runs in which 
dust is included have a ``D'' appended. To give an example: 
run I25-M3D is a run performed at redshift $z=25$ with an 
intermediate-sized halo that has a metallicity ${\rm Z} = 10^{-2} \: {Z_{\odot}}$, an ultraviolet background field strength of $J_{21} = 0.1$,
and that includes the effects of dust.

\section{Results}

\subsection{Zero-Metallicity Gas}
\label{sec:zero}
Before examining the behaviour of the metal-enriched gas, it is prudent to
first examine what happens in the absence of metals, i.e.\ in gas of wholly
primordial composition. Previous work suggests that there is
a critical protogalactic mass $M_{\rm crit}$ such that gas in protogalaxies 
with $M > M_{\rm crit}$ is able to cool and collapse within a Hubble time,
while gas in protogalaxies with $M < M_{\rm crit}$ cannot 
\citep[see e.g.][]{TEG97,MAC01,YOS03,MAS06,OSH06}.

\begin{figure}
\centering
\includegraphics[width=20pc,angle=-90]{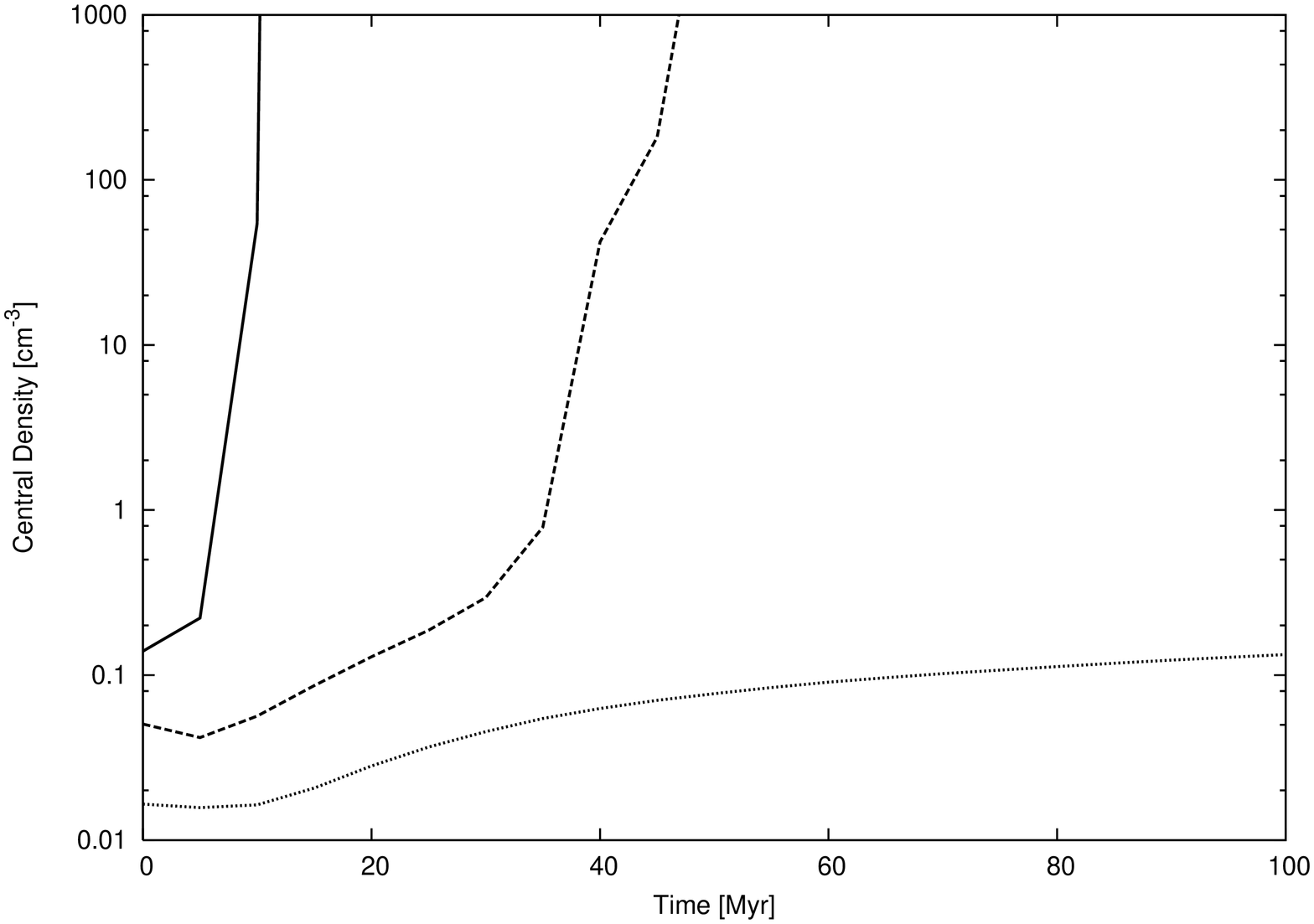}  
\includegraphics[width=20pc,angle=-90]{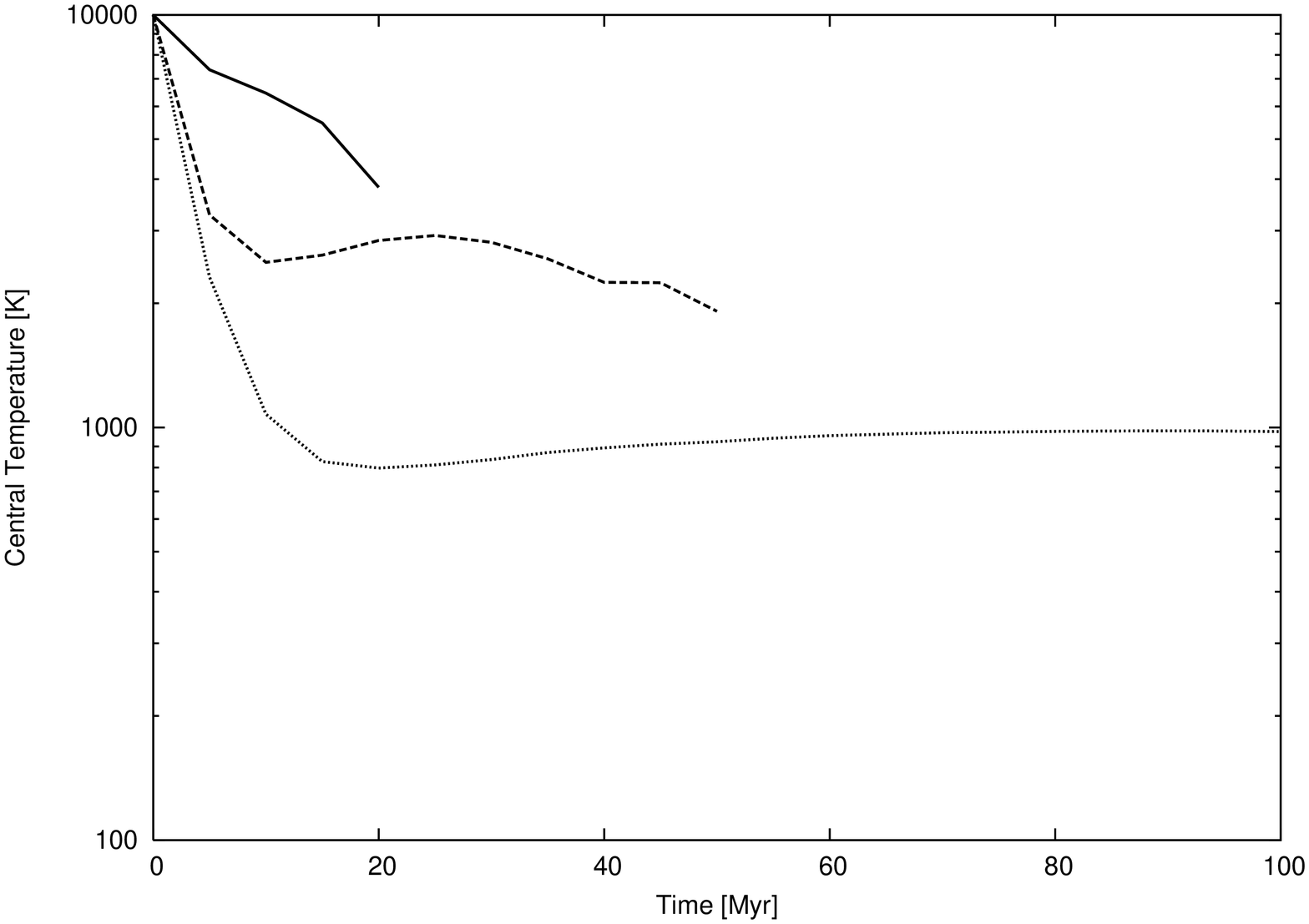}  
\caption{Top: Time evolution of the gas density within the scale radius
$r_{s}$ of the dark matter halo in runs L30-Z1 ({\it solid line}), 
I30-Z1 ({\it dashed line}), and S30-Z1 ({\it dotted line}), corresponding to halos with
masses $4.8 \times 10^{6}$, $4.8 \times 10^{5}$, and $4.0 \times 10^{4} \: \msun$, respectively.
Bottom: Same as the top panel, but for the central temperature of the gas.
\label{Mcrit}}
\end{figure}

To verify that 
we recover the same behaviour, we performed a set of three runs at $z=30$
with halo masses spanning a range of 2 orders of magnitude. The halos
in these three runs, which we designate as S30-Z1, I30-Z1, and L30-Z1, had 
masses of $4.0 \times 10^{4} \: \msun$, $4.8 \times 10^{5} \: \msun$ and 
$4.8 \times 10^{6} \: \msun$ respectively. In Figure~\ref{Mcrit}, we show 
the time evolution of the mean number density and mean temperature within the 
scale radius~$r_{{s}}$ in these runs. 

Table~\ref{tab:run_results} gives the properties of the gas within the scale
radius after $100\,\mathrm{Myr}$ of evolution. The table lists the minimum
central temperature~$T_{c, {\rm min}}$ and the maximum central density~$n_{c, {\rm max}}$. 
We also calculate the fractional $\mHt$ abundance with respect to the total number 
of hydrogen nuclei, and list the maximum central value, $x_{{\rm \mHt}, c,{\rm max}}$, 
in the table. 

The mass fraction of gas within a sink particle is denoted by 
$f_{\mathrm{sink}}$. Because we have no information on the density and temperature 
distribution within the sink particle radius, we adopt the values at time of 
sink particle formation in Table~\ref{tab:run_results}. Note that as the gas 
in run S30-Z1 does not collapse, it never reaches the density required for 
a sink particle to form.

Both Figure~\ref{Mcrit} and Table~\ref{tab:run_results} clearly 
demonstrate the change in behaviour that occurs for increasing $M$: 
gas in run L30-Z1 cools and collapses very quickly, on a timescale 
$t \ll t_{\rm H}$, while gas in run I30-Z1 cools and collapses on a 
slower timescale $t \sim t_{\rm H}$. Finally, the gas in run S30-Z1 
cools, but not by enough to bring its temperature below the halo 
virial temperature, and so it does not collapse within a Hubble time.
Indeed, we have run this simulation for $\sim 520 \: {\rm Myr}$ longer 
than is shown in the plot, but find no evidence for collapse even after 
this extended period. We can therefore conclude that $4.0 \times 10^{4}\: \msun
< M_{\rm crit} < 4.8 \times 10^{5} \: \msun$, with the value likely 
somewhere near the upper end of this range. For comparison, 
\citet{TEG97} find that $M_{\rm crit} = 10^{6} \: \msun$ at $z=30$; 
\citet{MAC01} find a significantly smaller value of 
$M_{\rm crit} = 1.25 \times 10^{5} \: \msun$ at the same redshift; 
and \citet{YOS03} find that $M_{\rm crit} \simeq 7.0 \times 10^{5} 
\: \msun$ at $z < 25$.  More recently, \citet{OSH06} have shown that
in their ensemble of simulations of primordial star formation,
the mass of the first star-forming halo lies in the range
$1.4 \times 10^{5} < M < 6.9 \times 10^{5} \: \msun$, where the 
scatter appears to be a result of the sensitivity of the problem to
the initial conditions. Our own estimate of $M_{\rm crit}$ is in
general agreement with these determinations, suggesting that our 
basic treatment of the physics of the primordial gas is sound.

\begin{figure}
\centering
\includegraphics[width=20pc,angle=-90]{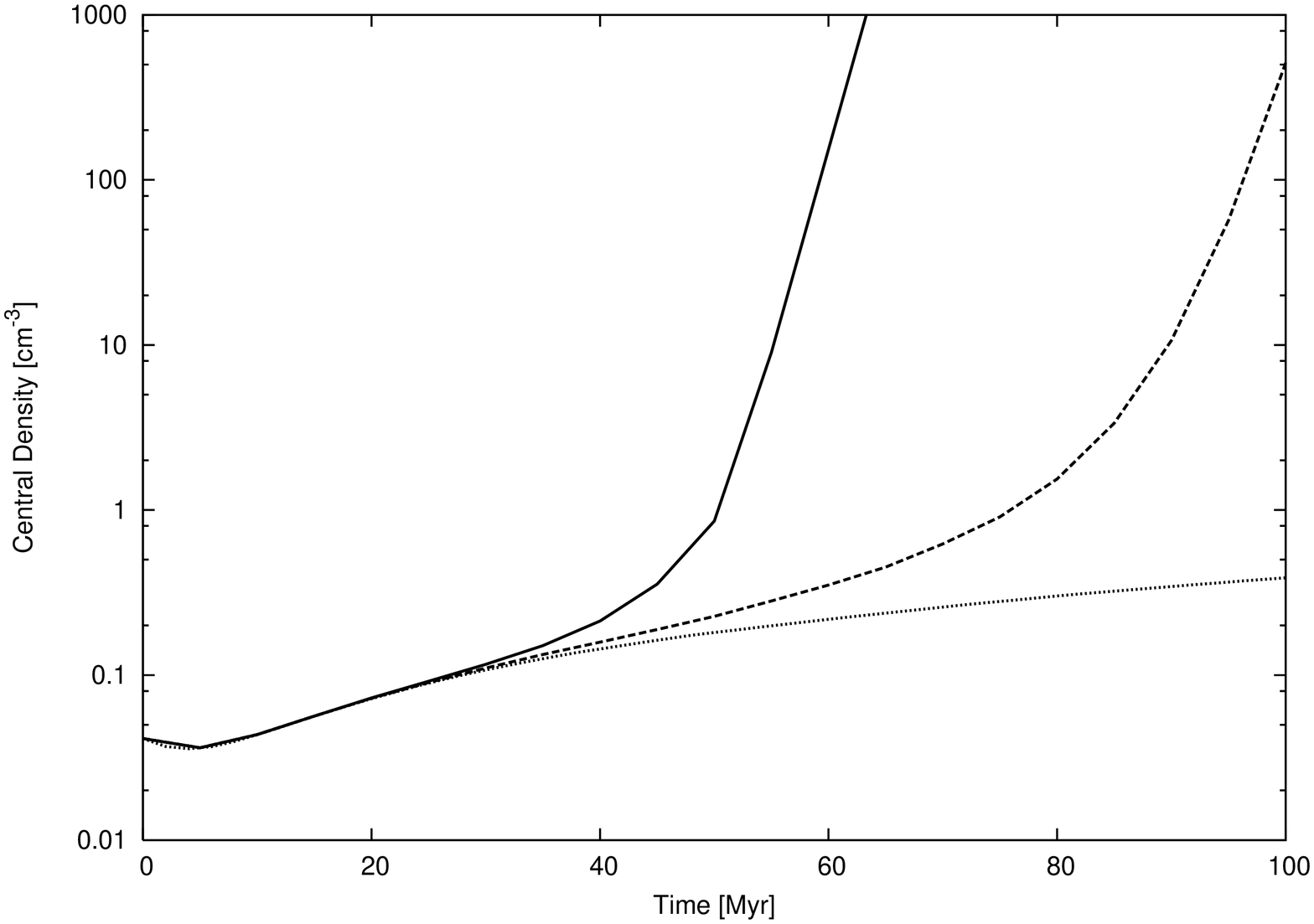}  
\includegraphics[width=20pc,angle=-90]{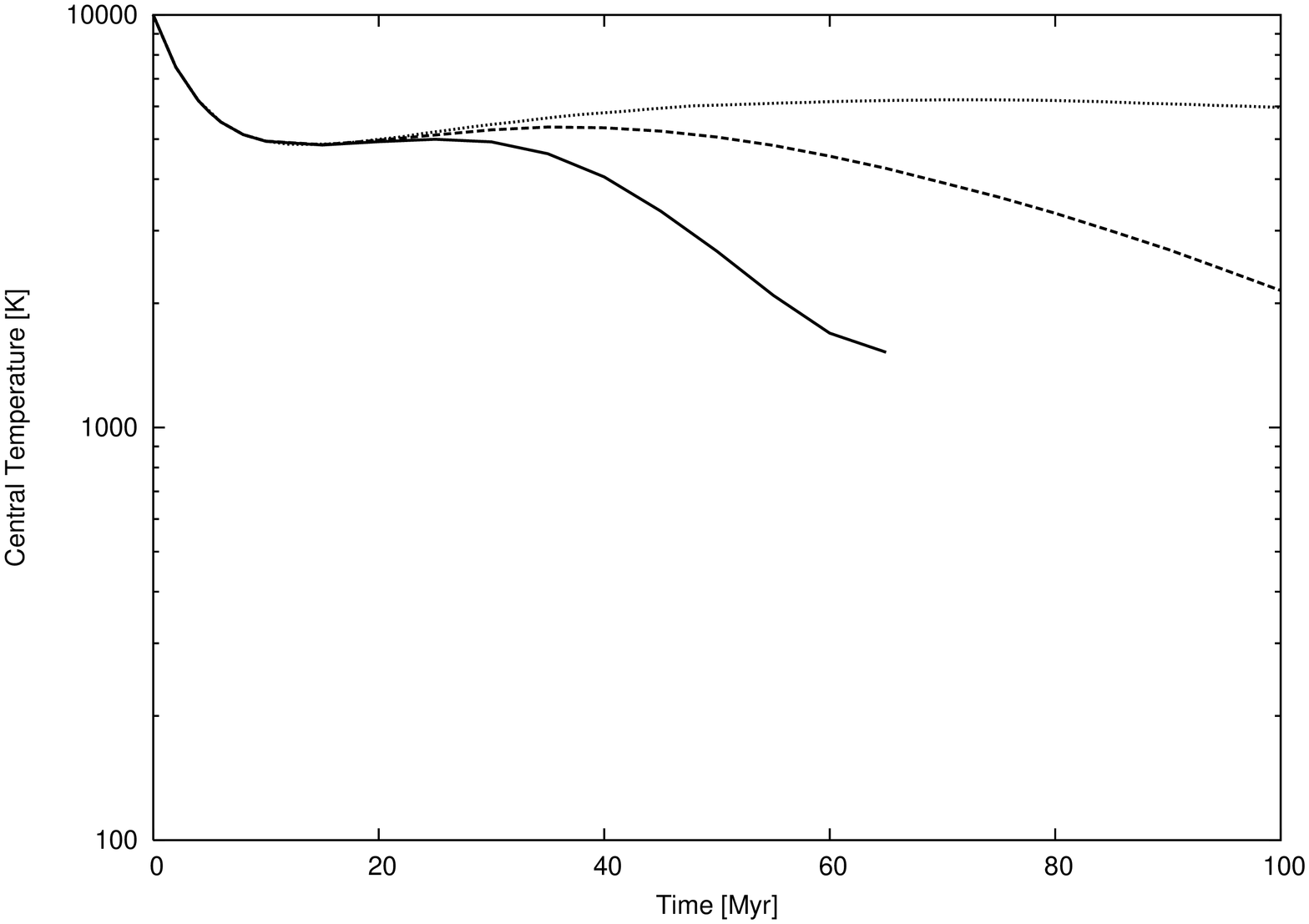}  
\caption{Top: Same as the top panel of Fig.~\ref{Mcrit}, but for runs I25-Z1 ({\it solid line}),
I25-Z2 ({\it dashed line}), and I25-Z3 ({\it dotted line}). These runs were started
at $z=25$ with a halo mass $M = 7.8 \times 10^{5} \: \msun$, and had
UV background field strengths $J_{21} = 0.0,\,0.01$, and $0.1$, respectively.
Bottom: Same as the top panel, but for the central temperature of the gas.
\label{Jcrit}}
\end{figure}

Another area that we can explore with our zero-metallicity runs is the
response of the gas to the presence of an ultraviolet background. 
Previous work has shown that $\mHt$ cooling of protogalactic gas 
is significantly affected once the strength of the background 
exceeds $J_{21} \simeq 0.01$ and that collapse in small halos is
almost entirely suppressed for $J_{21} \geq 0.1$, owing to
efficient photodissociation of the $\mHt$ \citep{HAI00,MAC01,YOS03,MAS06}. 
We show in Figure~\ref{Jcrit} an example of what happens in our own 
simulations as the strength of the background is increased. 
In the figure, we plot results from three runs performed at 
$z=25$ with halo mass $M = 7.8 \times 10^{5}\: \msun$. These runs are 
designated I25-Z1, I25-Z2, and I25-Z3 and had UV background field 
strengths of $J_{21} = 0.0, 0.01$ and 0.1, respectively. We see from 
the figure that the imposed UV background of strength $J_{21}=0.01$ 
in run I25-Z2 delays collapse by more than $30 \:\mathrm{Myr}$ 
compared to run I25-Z1, but that collapse does eventually occur.
On the other hand, a further increase in the UV background field
strength by a factor of 10 completely inhibits the collapse of 
the gas within the time frame of our simulation. These results 
are in good agreement with the previous studies cited above,
further increasing our confidence that our basic modelling is 
sound.

\subsection{Metal-Enriched Gas}
\label{high}
\subsubsection{Low-Metallicity Protogalaxies}
Having established how primordial gas behaves in these
conditions, we next examined the effect of enriching the
gas with metals. As previously mentioned in \S~\ref{IC},
we focussed initially on gas with a metallicity 
${\rm Z} = 10^{-3} \: {Z_{\odot}}$. This metallicity
is at the upper limit of the range of values proposed 
for the so-called critical metallicity ${Z_{\rm crit}}$,
the value of the metallicity at which efficient fragmentation 
and low-mass star formation is hypothesized to first occur 
\citep{BRO01,BL03,SCH02,OMU05}. It is also comparable to the 
globally averaged metallicity produced by the sources responsible 
for reionization \citep[see e.g.][]{RIC04}.

\begin{figure}
\centering
\includegraphics[width=25pc]{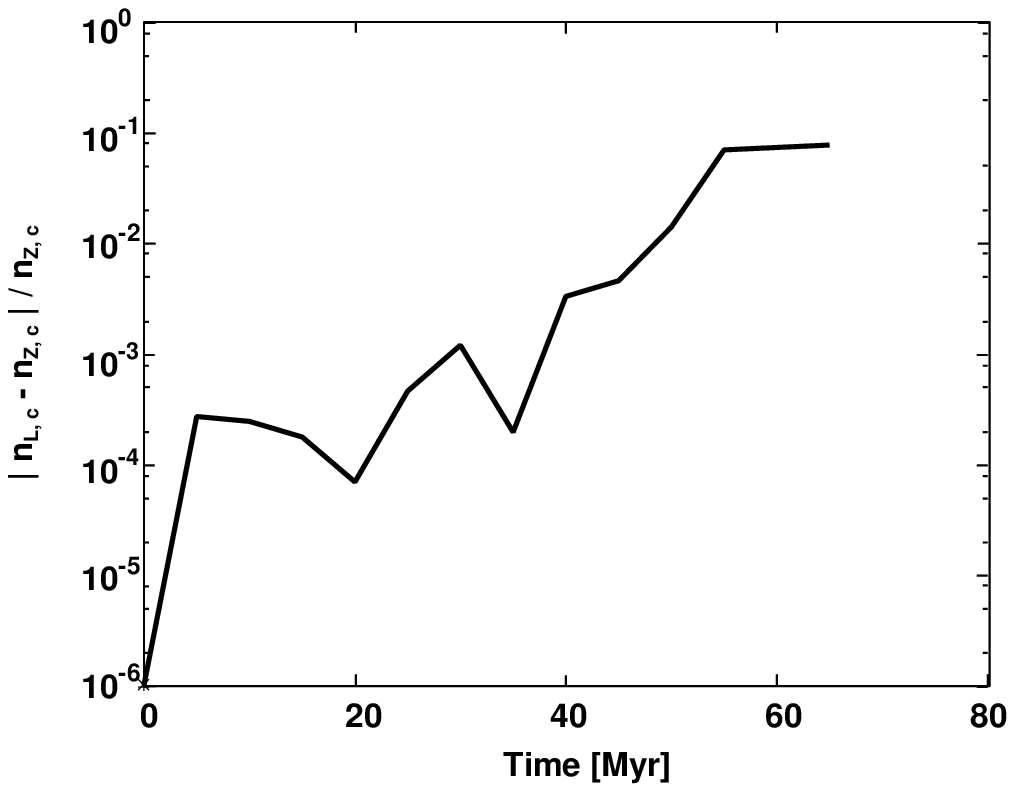}   
\includegraphics[width=25pc]{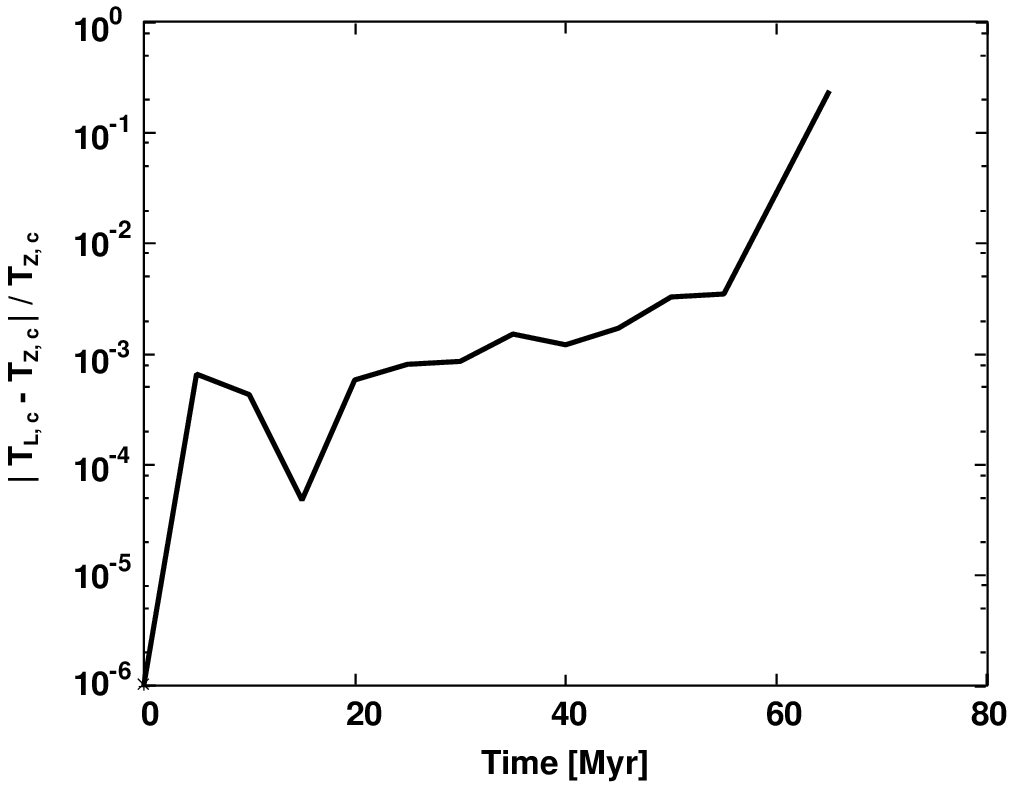}   
\caption{Top: Time evolution of the relative difference  
$|n_{\mathrm{Z}, c}-n_{\mathrm{L}, c}|/n_{\mathrm{Z}, c}$, where $n_{\mathrm{L}, c}$ is the 
 central gas density in run I25-L1 and $n_{\mathrm{Z}, c}$ the central density in
run I25-Z1. For clarity, we only plot the evolution until the point at which a 
sink particle forms (or until the end of the run, if no sink forms).
Bottom: Same as the top panel, but for the relative difference between the central temperature
in the two runs. \label{rel-diff}}
\end{figure}

In the top panel of Figure~\ref{rel-diff} we show the time evolution of the 
relative difference between the central density in run I25-L1 
and in I25-Z1. In other words, 
we plot $|n_{\mathrm{Z}, c}-n_{\mathrm{L}, c}| / 
n_{\mathrm{Z}, c}$, where $n_{\mathrm{L}, c}$ and $n_{\mathrm{Z}, c}$
are the central densities in runs I25-L1 and I25-Z1, respectively. 
We plot this quantity to highlight the difference between the 
two runs, which would be difficult to discern in a more conventional 
comparison plot. We see that  prior to sink particle formation,
the central density of the two runs differs by less than 10\%,
and that at early times ($t < 50 \: {\rm Myr}$), the difference
is less than 1\%. In the bottom panel of Figure~\ref{rel-diff} we show a similar 
plot for the central temperature in the two runs. Again, the 
difference between the two runs is very small.

These results demonstrate that the cooling of the gas at
low densities ($n < 1 \: {\rm cm^{-3}}$) and high 
temperatures ($T > 2000 \: {\rm K}$) is barely influenced 
by the presence of metals: fine-structure cooling contributes 
only marginally to the total cooling rate, and the results
of the runs agree to within 1\%. The evolution of 
cooling during the first $50\,\mathrm{Myr}$ and the onset of 
collapse are thus almost independent of the metallicity of 
the gas, at least for gas with ${Z} \leq 10^{-3} \: 
{Z_{\odot}}$. At later times during the collapse, the
difference between the metal-enriched and the zero-metallicity 
models becomes more pronounced, indicating that metal-line
cooling is beginning to play a more significant role.
Despite this, the behaviour of the gas in the two runs
remains very similar up to the point at which sink
formation occurs (i.e.\ at $n \leq 500 \: {\rm cm^{-3}}$).

An obvious question to ask is whether these results are
somehow peculiar to the particular combination of halo
mass and redshift that we have chosen to examine. To 
investigate this, we performed additional simulations
with both ${Z} = 0$ and $10^{-3} \: 
{Z_{\odot}}$ for a range of different halo masses 
and redshifts, as summarized in Table~\ref{tab:run_params}. 
We found that in all cases we obtained similar results; 
the thermal and dynamical evolution of gas in runs with
${Z} = 10^{-3} \: {Z_{\odot}}$ differed very 
little from that of gas in runs with ${Z} = 0$.

\subsubsection{The effects of rotation}
We have also investigated the possibility that our results
were significantly affected by our neglect of the effects 
of rotation in the majority of our runs. Theoretically, 
we do not expect rotation to be important on scales 
$r \gg \lambda r_{\rm vir}$, where $\lambda$ is the
dimensionless spin parameter, given by \citep{peeb71}
\begin{equation}
\lambda = \frac{J |E|^{1/2}}{G M^{5/2}},
\end{equation}
where $J$ is the total angular momentum of the halo,
$M$ is the halo mass and $E$ is the total (kinetic
plus potential) energy of the halo. Previous work 
has shown that for halos with the range of masses and 
redshifts considered in this paper, $\lambda$ has a
lognormal distribution, with mean $\bar{\lambda} = 
0.035$ \citep{YOS03}. Therefore, we expect that 
rotation will only significantly affect the dynamics
of the gas within the innermost few percent of the 
halo. As this is typically comparable to the radius
at which we cease to be able to resolve the collapse
of the gas, we do not expect the inclusion of 
rotation to significantly affect our results or to
materially alter our conclusions.

\begin{figure}
\centering
\includegraphics[width=20pc,angle=-90]{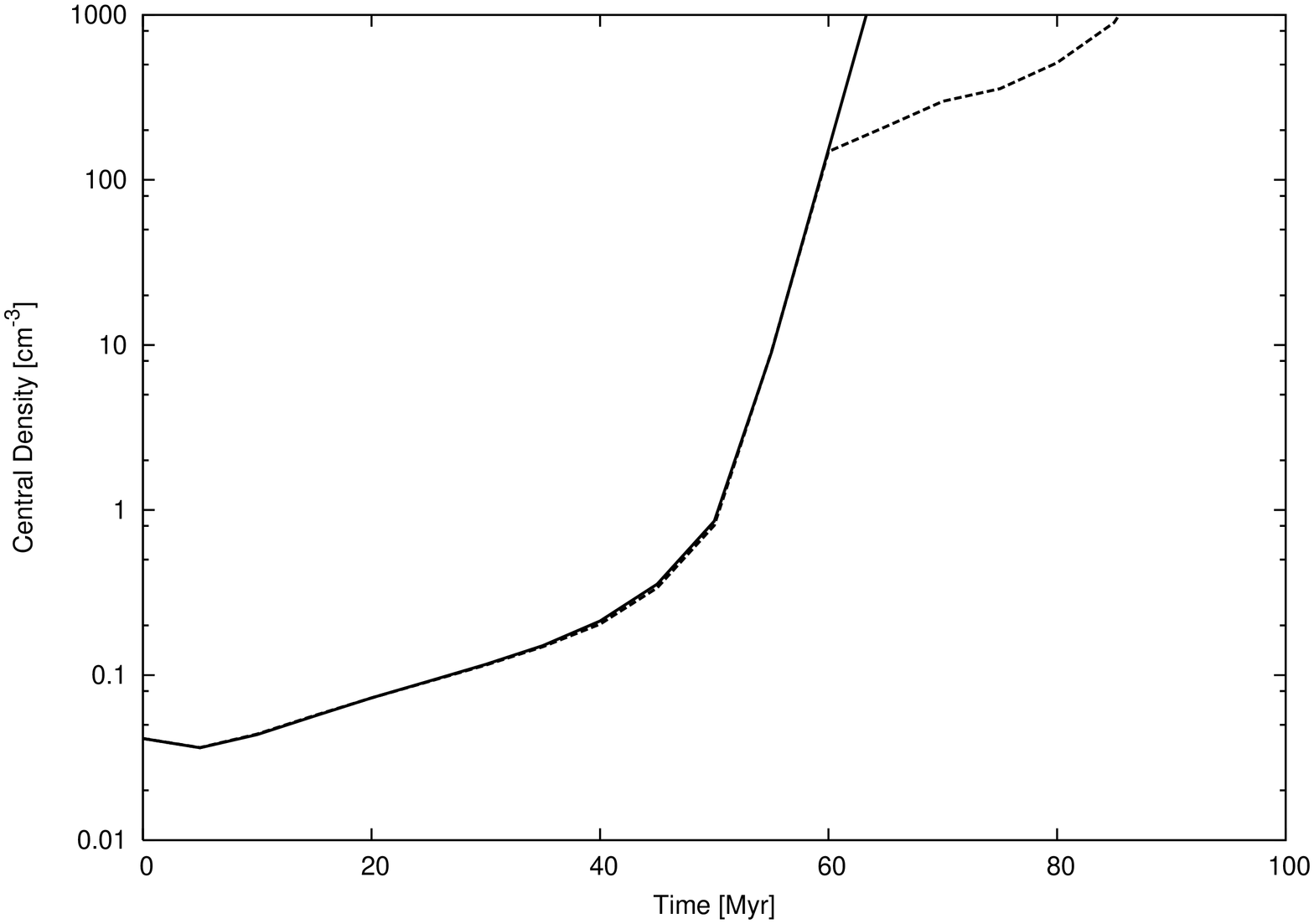} 
\includegraphics[width=20pc,angle=-90]{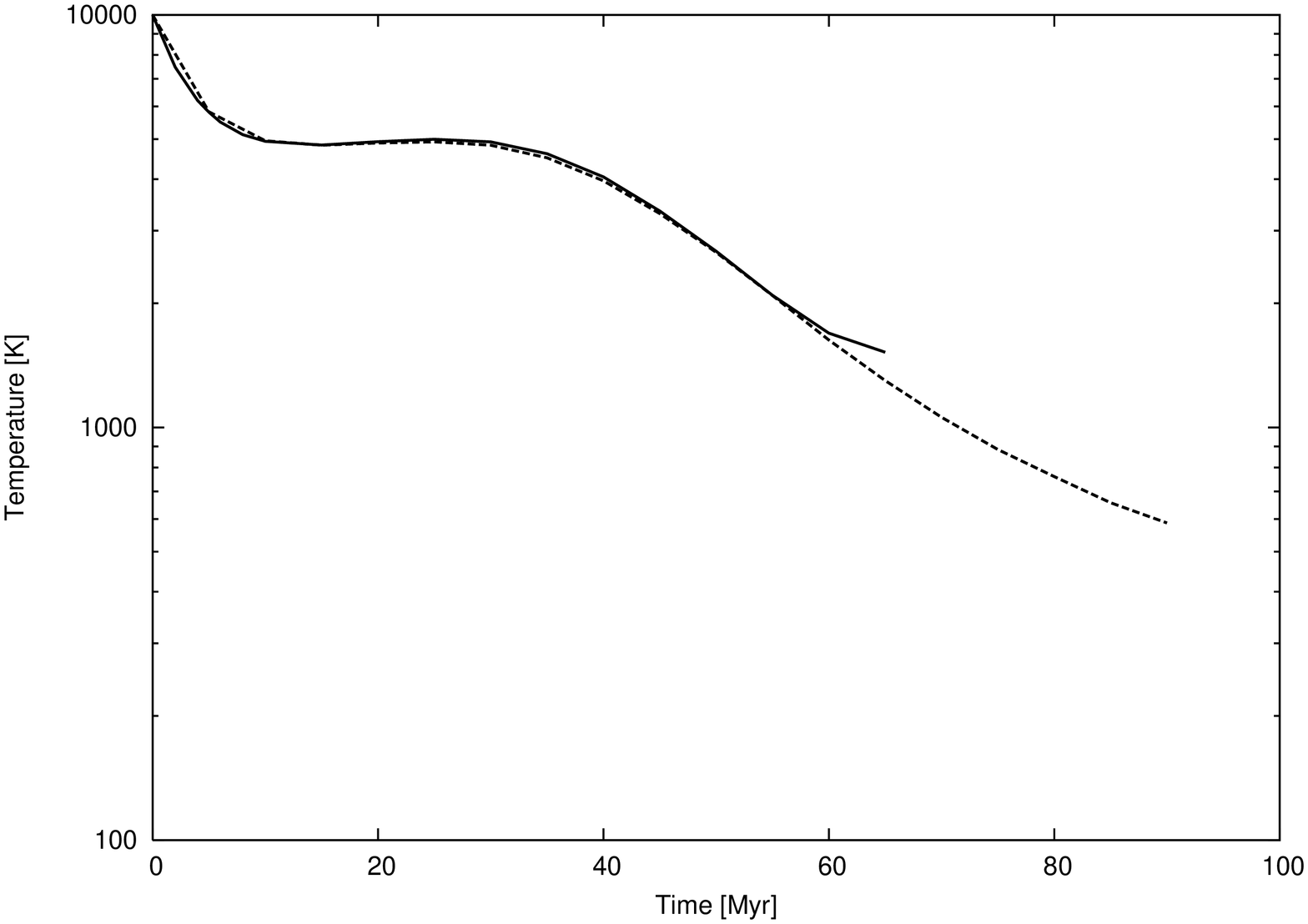} 
\caption{Top: Time evolution of the gas density within the scale radius
$r_{s}$ of the dark matter halo in runs I25-Z1 ({\it solid line}) and
I25-Z1-ROT ({\it dashed line}). 
Bottom: Same as the top panel, but for the central temperature of the gas \label{rot-plot}}
\end{figure}

In order to verify that this is a reasonable expectation, 
we performed one simulation in which the initial gas
distribution was given a non-zero angular momentum.
Within the virial radius of the halo, the gas was placed
into rotation with constant angular velocity. At larger
radii, the initial angular velocity decreased linearly
with radius, reaching zero at the truncation radius of 
the halo. This run, which we designate hereafter as 
I25-Z1-ROT, had a spin parameter $\lambda = 0.05$. 
The other parameters used for this run were the same as
in run I25-Z1. In Figure~\ref{rot-plot}, we compare the 
outcome of these two simulations. We see that, as expected, 
rotation has very little effect on the evolution of the 
gas at early times. It does appear to affect the evolution
once the density exceeds $n = 100 \: {\rm cm^{-3}}$,
significantly slowing the collapse, but at this point we 
are close to the resolution limit of our simulation
and so we are unable to say with any certainty whether 
or not a rotationally supported disk forms. Nevertheless,
as the main results of this paper concern the evolution
of the gas at densities $n \ll 100 \: {\rm cm^{-3}}$, it
is clear that they are unaffected by our neglect of 
rotation in the majority of our simulations.

\subsubsection{When {\em Do} Metals Make a Difference?}
\begin{figure}
\centering
\includegraphics[width=20pc,angle=-90]{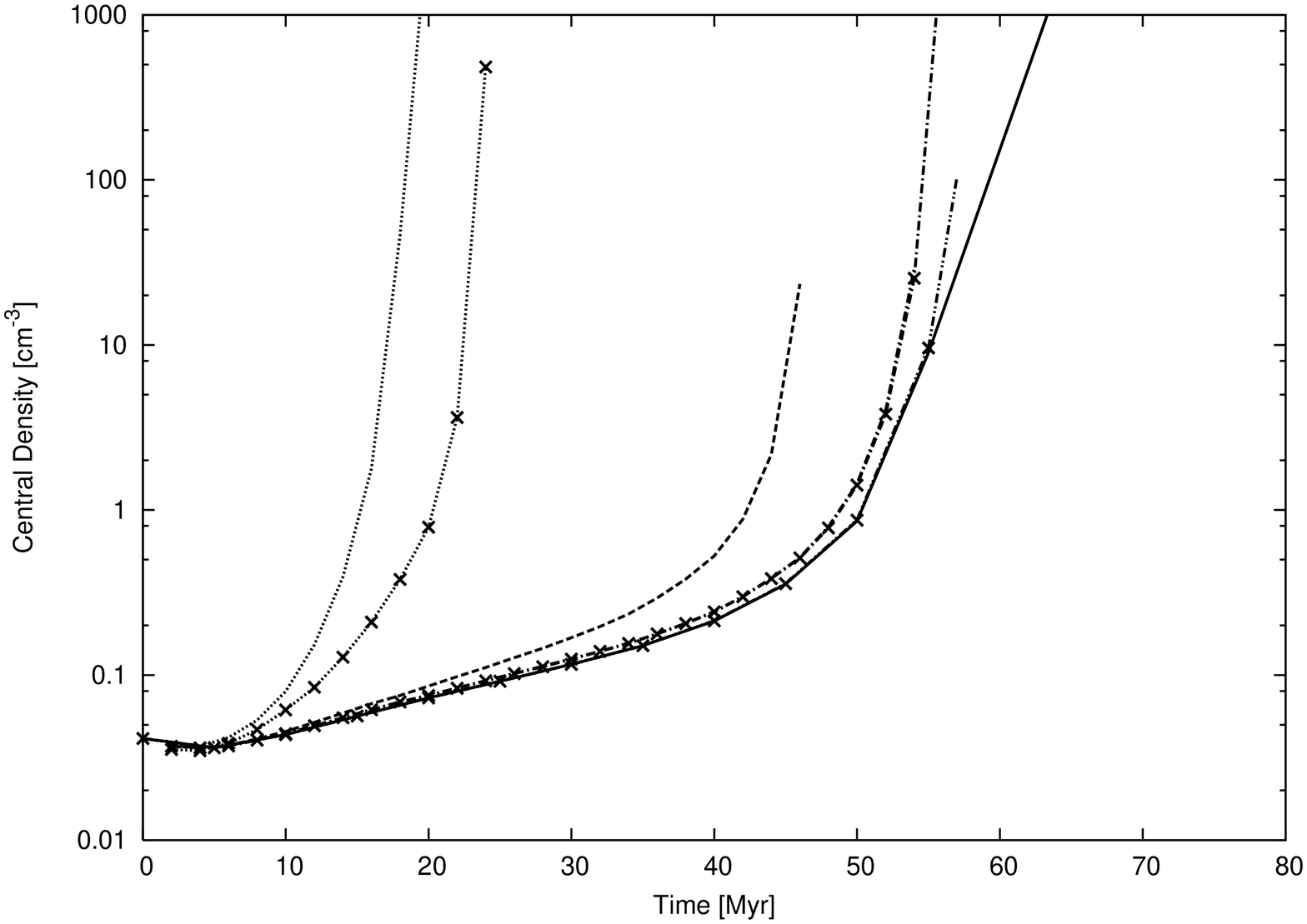} 
\includegraphics[width=20pc,angle=-90]{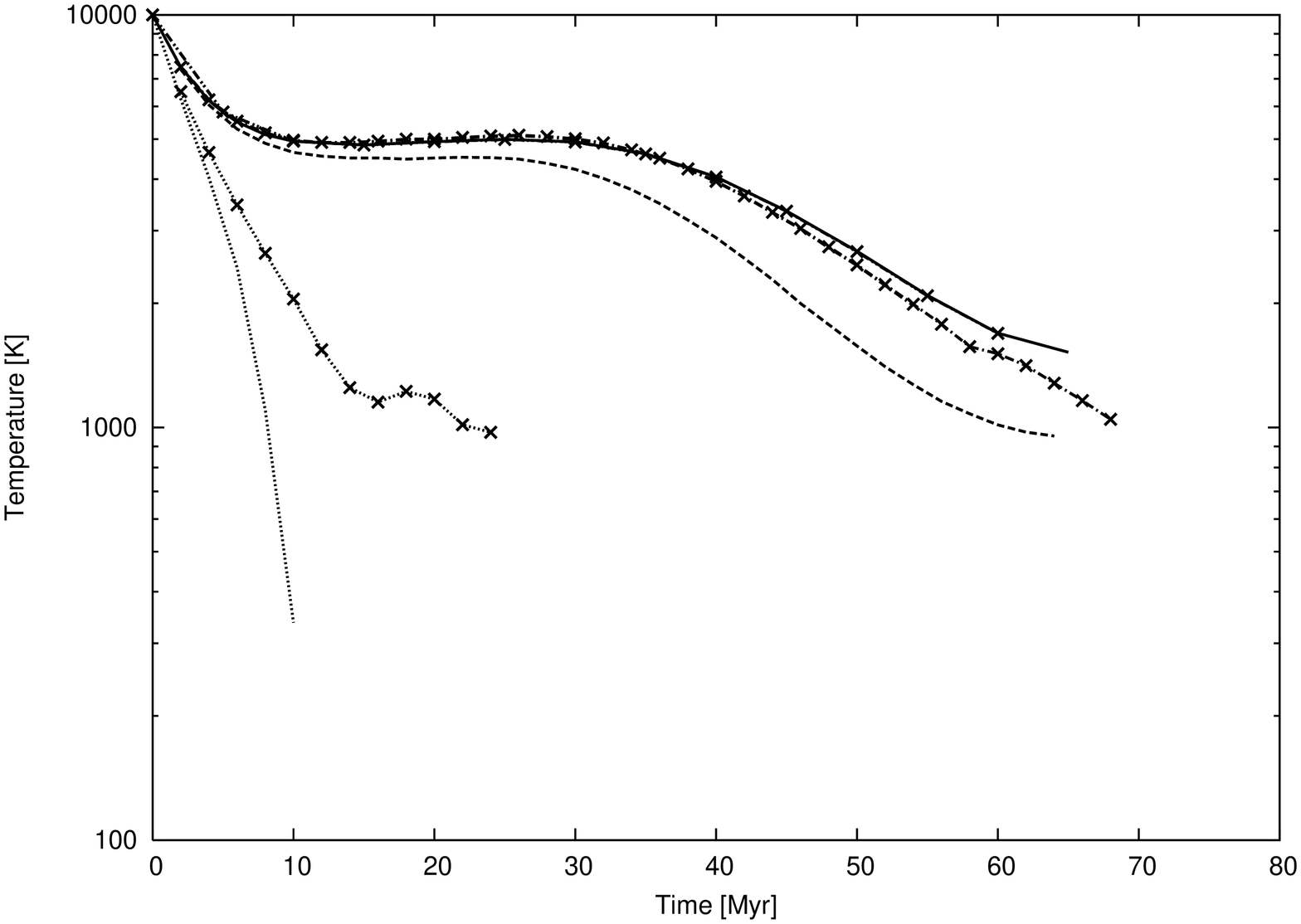} 
\caption{Top: Time evolution of the gas density within the scale radius
$r_{\rm s}$ of the dark matter halo in runs I25-Z1 ({\it solid line}),
I25-L1 ({\it dot-dot-dashed line}), I25-M1 ({\it dash-dotted line}),
I25-N1 ({\it dashed line}), and I25-S1 ({\it dotted line}), which have 
metallicities ${Z} = 0.0,\,10^{-3},\,10^{-2},\,0.1$, and 
$1.0 \: {Z_{\odot}}$, respectively. We also include three runs 
with dust, I25-L1D, I25-M1D, and I25-S1D, denoted by the lines with 
symbols. In all of these runs, the halo mass $M = 7.8 \times 10^{5} 
\: \msun$ and the run begins at a redshift $z=25$. 
No ultraviolet background is present in any of these runs.
Bottom: Same as the top panel, but for the central temperature of the gas.
\label{Z-comp}}
\end{figure}

If cooling from metals is ineffective when ${Z} = 
10^{-3} \: {Z_{\odot}}$, then just when does it 
become important? How enriched does the gas need to
be before metal-line cooling makes a significant 
difference? To investigate this, we performed a set
of additional runs using our fiducial halo parameters 
($z=25$, $M = 7.8 \times 10^{5} \: \msun$) that 
looked at the effects of further increasing ${Z}$.
In Figure~\ref{Z-comp}, we plot the time evolution of 
the central density and central temperature
in runs I25-Z1, I25-L1, I25-M1, I25-N1, and I25-S1, 
which had metallicities of ${Z} = 0.0,\,10^{-3},\, 
10^{-2},\,0.1$ and $1.0 \: {Z_{\odot}}$, respectively.
We also show in the plot the results of runs I25-L1D, 
I25-M1D, and I25-S1D, which had metallicities of
${\rm Z} = 10^{-3},\,10^{-2}$ and $1.0 \: {\rm Z_{\odot}}$,
respectively, and which differed from runs I25-L1,
I25-M1, and I25-S1 by including the chemical and 
thermal effects of dust.

It is clear from the figure that all of the runs with
${Z} \leq 0.01 \: {Z_{\odot}}$ evolve in essentially
the same way. Some minor differences in the rate of
collapse and in the central temperature of the gas are
apparent at late times, particularly in the two runs
with ${Z} = 0.01 \: {Z_{\odot}}$, but these do
not appear likely to significantly affect the outcome of
the simulation. If we enrich the gas by another order
of magnitude, to ${Z} = 0.1 \: {Z_{\odot}}$, 
however, then we begin to see the metals playing a 
significant role in the cooling of the gas. In run
I25-N1 the gas cools more rapidly than in any of the
lower metallicity runs, and as a consequence, runaway
gravitational collapse sets in approximately 
$10 \: {\rm Myr}$ earlier than in the lower metallicity 
runs. Nevertheless, the evolution of the gas in this run
remains qualitatively very similar to that in the lower
metallicity runs. It is only when we enrich the gas by
a further order of magnitude, to solar metallicity, 
that we see a truly dramatic difference in the outcome
of the simulation, with the gas in runs I25-S1 and 
I25-S1D cooling and collapsing within the first 
$20 \: {\rm Myr}$ of the run. Based on these results,
we would estimate the critical metallicity at which
metal cooling dominates in the low-density gas to 
be ${Z_{\rm crit}} \sim 0.1 \: {Z_{\odot}}$, to
the nearest order of magnitude.

Figure~\ref{Z-comp} also allows us to assess the role
played by dust in these systems. The fact that runs
I25-L1D and I25-M1D produce essentially identical 
results to runs I25-L1 and I25-M1 demonstrates that
at ${Z} = 0.01 \: {Z_{\odot}}$ and below, 
dust has a minimal effect on the low-density evolution
of these systems. On the other hand, at high 
metallicities, dust {\em does} significantly affect 
the evolution of the gas. Although the presence of
dust might be expected to aid the cooling of the 
gas, it is clear from Figure~\ref{Z-comp} that in
actual fact the gas cools more rapidly when dust 
is not present. This appears to be a consequence of
the fact that the ionized gas recombines more rapidly
when dust is present, owing to recombination of 
ionized hydrogen on the surface of dust grains, 
with the result that both the Compton cooling 
rate, as well as the contributions of electron-ion 
and electron-atom collisions to the fine-structure 
cooling rate, fall off more rapidly with time.

\subsubsection{Collapse with a UV background}
\begin{figure}
\centering
\includegraphics[width=20pc,angle=-90]{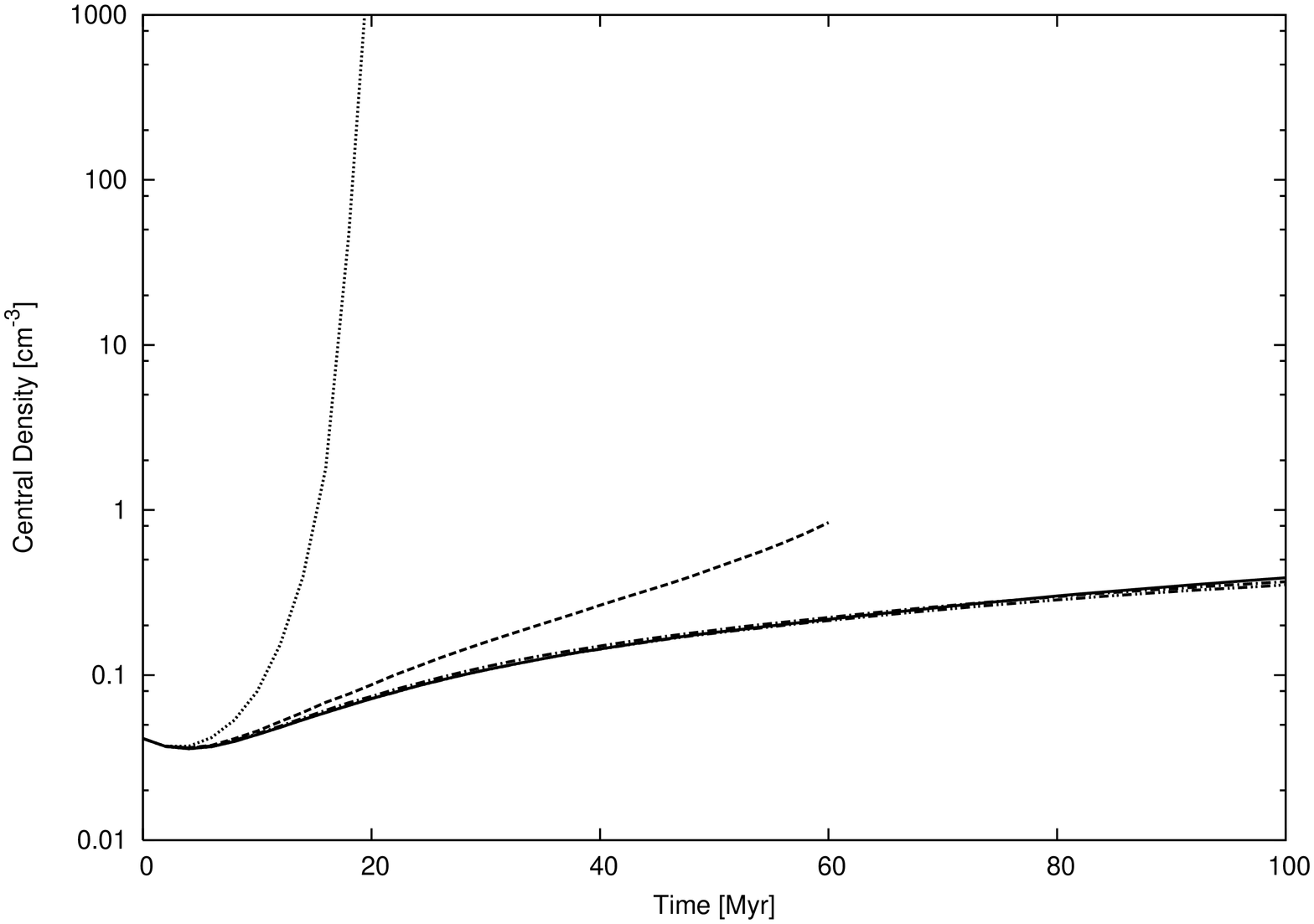} 
\includegraphics[width=20pc,angle=-90]{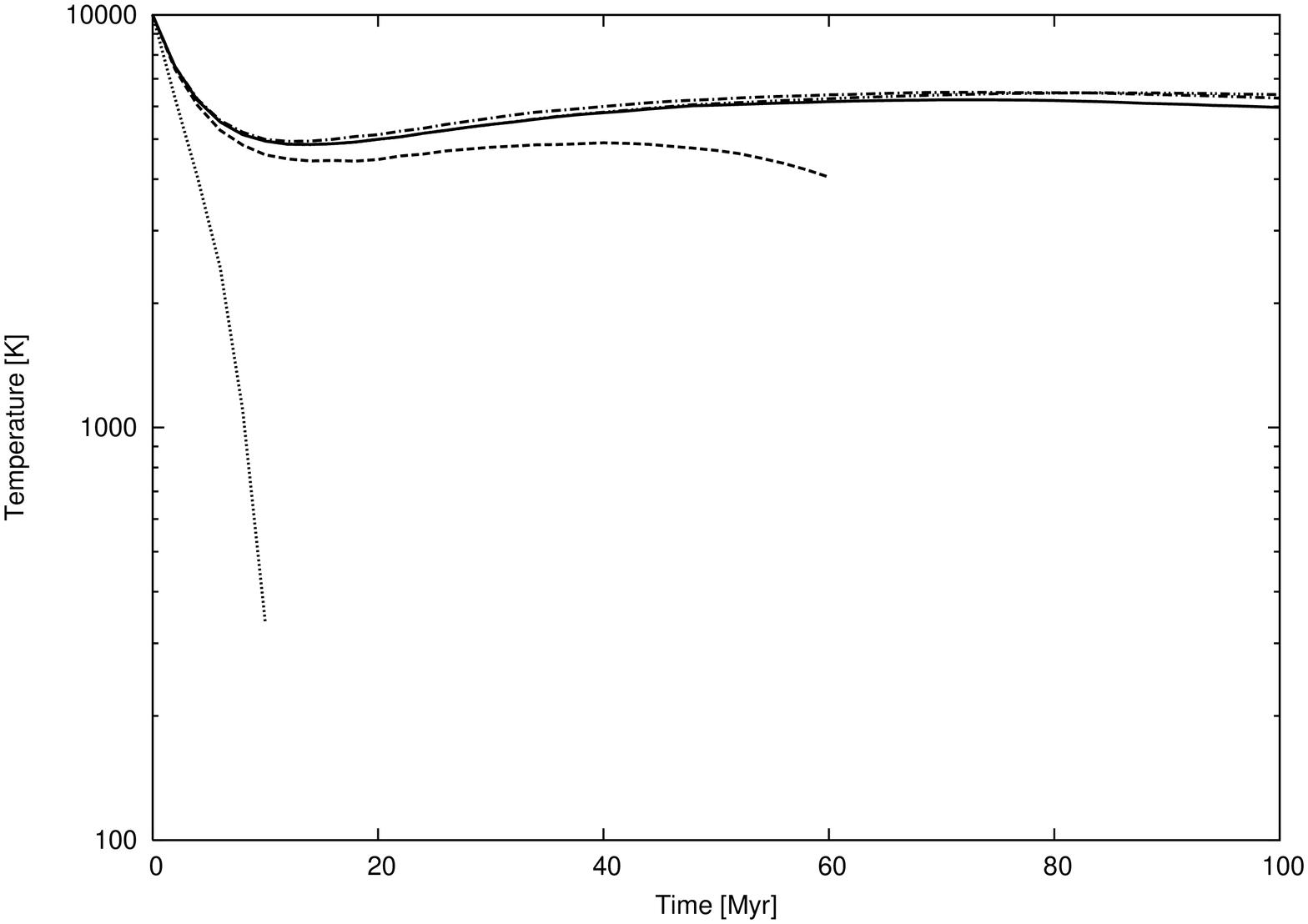} 
\caption{Top: Same as Fig.~\ref{Z-comp}, but for runs I25-Z3 ({\it solid line}),
I25-L3 ({\it dot-dot-dashed line}), I25-M3 ({\it dash-dotted line}), 
I25-N3 ({\it dashed line}), and I25-S3 ({\it dotted line}). The UV background field 
strength in all of these runs was $J_{21} = 0.1$.
Bottom: Same as the top panel, but for the central temperature of the gas.
\label{Jcrit-Z}}
\end{figure}

Finally, it is important to determine whether
metal-line cooling allows collapse to occur in 
UV-irradiated halos that would otherwise be unable 
to collapse. Figure~\ref{Jcrit-Z} 
summarizes the results of a set of simulations designed 
to address this question. It shows the evolution of the
central density and temperature of the gas in a set of
runs in which $J_{21} = 0.1$, $M$ and $z$ were held constant
at their fiducial values, and the metallicity was varied. 
Details of the eventual state of the gas at the end of 
these runs are given in Table~\ref{tab:run_results}. 
We see that even at metallicities 
as high as ${Z} = 0.01 \: {Z_{\odot}}$, an ultraviolet
background flux of strength $J_{21} = 0.1$ is sufficient  
to completely suppress collapse. We therefore conclude
that {\em in low-metallicity, low-mass protogalactic halos, 
metal-line cooling cannot initiate collapse}. Either the
surviving $\mHt$ provides enough cooling to allow collapse 
to occur, irrespective of whether or not there are metals
present, or the gas does {\em not} collapse. Only at quite high
metallicities (${Z} \geq 0.1 \: {Z_{\odot}}$) do the
metals actually make an appreciable difference to the
evolution of the gas.

\section{Discussion}
At first sight, the fact that fine structure cooling from metals has little impact on 
the thermal or dynamical evolution of the gas at  metallicities below
$0.1 \: {Z_{\odot}}$ is somewhat surprising, given that \citet{BRO01} found 
that gas with a metallicity of only ${Z} = 10^{-3} \: {Z_{\odot}}$ could cool 
rapidly and fragment even in the complete absence of molecular hydrogen. 
However, comparison of the cooling time due to fine-structure emission with the 
free-fall time of the gas helps to make the situation clear. The cooling time 
due to fine-structure emission is given by
\begin{equation}
t_{\rm cool, fs} = \frac{1}{\gamma - 1} \frac{n k T}{\Lambda_{\rm fs}}, 
\end{equation}
where $\Lambda_{\rm fs}$ is the total cooling rate per unit volume due to fine-structure emission, $T$ is the temperature of the gas, $\gamma$ is the
adiabatic index, $n$ is the number density of the gas, and $k$ is the Boltzmann
constant. The free-fall time can be written as
\begin{equation}
t_{\rm ff} = \left( \frac{3 \pi}{32 G \rho} \right)^{1/2},
\end{equation}
where $G$ is the gravitational constant and $\rho = \rho_{\rm gas} + \rho_{\rm
  dm}$. The gas density $\rho_{\rm gas}$ can be written in terms of the number
  density $n$ as $\rho_{\rm gas} = \mu n$, where $\mu$ is the mean molecular weight of the gas.

\begin{figure}
\centering
\includegraphics[width=25pc]{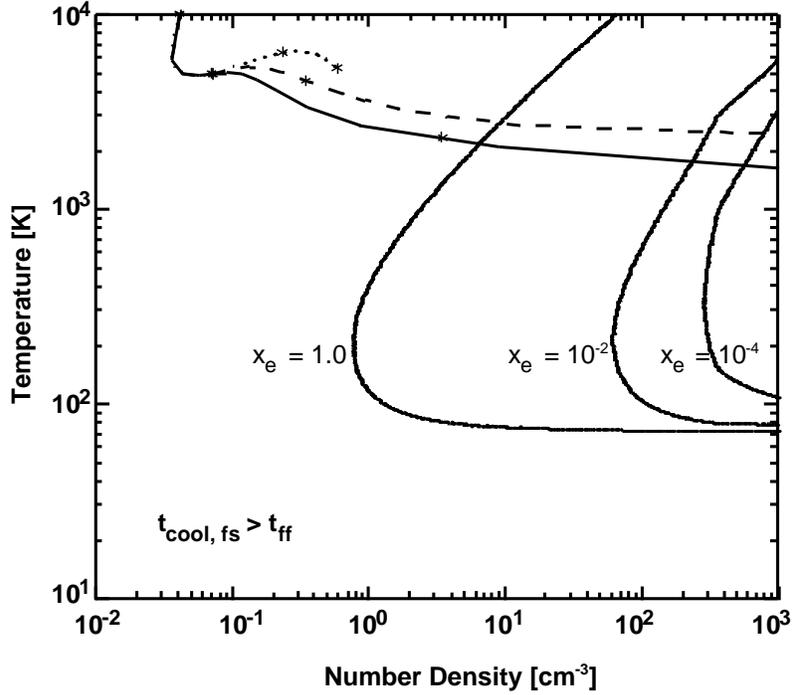}  
\caption{{\it Solid contours}: Temperature and density at which the cooling time
due to fine-structure emission, $t_{\rm cool, fs}$, equals the free-fall time, $t_{\rm ff}$,
for gas with a metallicity  ${Z} = 10^{-3}\, {Z}_{\odot}$ and with fractional ionizations 
$x_{\rm{e}} = 1.0$,  $10^{-2}$, and $10^{-4}$. To the
left of each line, $t_{\rm cool, fs} > t_{\rm ff}$, so metal cooling is inefficient.
In every case, we assume that all of the carbon and silicon is present as $\cp$ and $\sip$, respectively. In the $x_{\rm{e}} = 1.0$ case, we assume that all of the oxygen
is present in the form of $\op$, but otherwise that it is all $\mO$. To
compute the free-fall time, we take the density to be the sum of the gas
density $\rho_{g}$ and the dark matter density at the center of the halo
$\rho_{\rm dm}$. The figure also shows how the temperature and density of the gas at
the centre of the halo evolve in runs I25-L1 ({\it solid line}), I25-L2 ({\it dashed
line}), and I25-L3 ({\it dotted line}). The evolution of the fractional ionization
in these runs is indicated by the star symbols: $x_{\rm{e}} = 1.0$ for $T =
10^{4} \: {\rm K}$, and decreases by a factor of 10 between each successive star.}  
\label{fig:ratio}
\end{figure}

In Figure~\ref{fig:ratio}, we indicate the temperatures and densities at which 
$t_{\rm cool, fs} = t_{\rm ff}$ for $10^{-3}\, {Z}_{\odot}$ gas, for three 
different assumed fractional ionizations, with $x_{\rm{e}} = 1.0$  (fully ionized gas), 
$10^{-2}$, and $10^{-4}$. In each 
case, we assume that the carbon and silicon are present only as $\cp$ or $\sip$, since 
a fairly small external UV flux is sufficient to achieve this. In the
$x_{\rm{e}} = 1.0$ case, we also assume that all of the oxygen is $\op$, since 
charge transfer between oxygen and hydrogen, which have nearly identical 
ionization potentials, ensures that
$x_{\rm O^{+}} / x_{\rm O} \simeq x_{\rm H^{+}} / x_{\rm H}$. For the dark matter density
$\rho_{\rm dm}$ we adopt the value found at the center of our fiducial protogalactic halo at t=0,
$\rho_{\rm dm} \simeq 8 \times 10^{-22} \: {\rm g} \: {\rm cm^{-3}}$. In the plot, regions to
the left of the line have $t_{\rm cool, fs} > t_{\rm ff}$, while those to the right have 
$t_{\rm cool, fs} < t_{\rm ff}$.

At the beginning of our fiducial simulation, gas in the center of the halo has a temperature
$T = 10^{4} \: {\rm K}$ and a number density $n = 0.03\,\mathrm{cm}^{-3}$. 
It therefore lies outside of the regime
where fine-structure cooling is efficient, and so it is not surprising that we find that metal-line 
cooling is initially unimportant. As the gas cools, whether through Compton cooling
or $\mHt$ emission, and begins to compress as it falls into the halo, it moves towards 
the temperature and density regime in which fine-structure cooling is
effective. As an example, we plot in Figure~\ref{fig:ratio} the trajectory in the 
density-temperature plane followed by gas at the center of the halo in run I25-L1.
However, at the same time as the gas is cooling and collapsing, it is also recombining.
This causes the boundary of the efficient cooling regime to move to the right of the plot,
towards higher densities. The physical reason for this shift is the
fact that free electrons are much more effective than neutral hydrogen at exciting the 
${\rm C^{+}}$ and ${\rm Si^{+}}$ fine-structure lines, and so electron excitation dominates
for $x_{\rm{e}} > 10^{-2}$--$10^{-3}$, depending on the temperature.  The net effect is that 
fine-structure cooling remains of little importance until the gas is near the high-cooling 
regime. This only happens after considerable cooling and compression has already
taken place, and therefore does not occur at all if $\mHt$ cooling is ineffective, as is 
the case, for instance, in runs I25-Z3 or I25-L3. 

For gas with a metallicity greater than $10^{-3} \: {Z}_{\odot}$, fine-structure cooling
is effective in a larger region of the density-temperature plane. At low densities 
($n < 1 \: {\rm cm^{-3}}$), an increase in ${Z}$ of an order of magnitude shifts 
the solid lines in 
Figure~\ref{fig:ratio} by an order of magnitude with respect to density. This allows 
us to estimate the metallicity at which fine-structure cooling will dominate even in
the initial low-density gas. We find that fine-structure cooling should dominate for
metallicities in excess of $0.1$--$1.0 \: {Z}_{\odot}$, depending on the details 
of the evolutionary trajectory followed by the gas at early times. This estimate agrees
well with the results we obtain from our SPH simulations.

Changes to the halo mass or redshift of collapse will affect this estimate in two ways.
First, increasing the redshift of collapse will make the dark matter halo denser and
will therefore decrease $t_{\rm ff}$ at early times, when dark matter is gravitationally
dominant. This will shrink the region in which fine-structure cooling can effectively
operate. Second, altering the halo mass and redshift will alter the evolutionary 
trajectory followed by the gas. Specifically, since the initial temperature of the gas
is fixed at $10^{4} \: {\rm K}$ in these simulations, changes in $M$ or $z$ that 
increase the virial temperature of the halo will make the gas more gravitationally
unstable, allowing it to collapse more quickly. The collapsing gas will therefore
be hotter than in our fiducial case (as it will have had less time to cool), and will
also be slightly more ionized. On the other hand, if the virial temperature of the 
halo is lowered, then the gas will be more stable, will collapse more slowly, and
will therefore be cooler and less ionized than in our fiducial case. However,
in either case, we would not expect to see a change of more than a factor of a few
in the critical metallicity required for effective fine-structure cooling at low 
densities, and our estimate should therefore still be of the correct order of 
magnitude. 

Given these results, we next ask whether they can possibly be consistent
with the results of \citet{BRO01}, who find that in their simulations, fine
structure cooling is effective even for $Z = 10^{-3} \: {Z_{\odot}}$, which is
at least 2 orders of magnitude smaller than our value. In fact, their 
results do actually appear to be consistent with ours, despite the apparent
disagreement. The reason is due to the difference in the initial conditions
used for their simulations and for ours. The gas in their simulations has an 
initial temperature of 200~K at $z = 100$, and its temperature falls further 
due to adiabatic cooling in the IGM prior to the formation of their simulated 
protogalactic halo at $z\sim 30$. Since this halo has a mass of 
$2 \times 10^{6} \: {M}_{\odot}$ and a virial temperature 
$T_{\rm vir} \simeq 5000 \: {\rm K}$, the temperature of the gas at the 
moment at which the halo forms is very much smaller than the halo virial 
temperature. Consequently, thermal pressure is initially unable to prevent 
the collapse of gas into the halo. As it collapses, the gas heats up, and the 
thermal pressure eventually becomes large enough to halt the collapse.
However, this does not occur until the gas temperature is approximately equal
to the halo virial temperature, by which time the gas density has increased to
a value $n \simeq 300 \: {\rm cm^{-3}}$. As can be seen from 
Figure~\ref{fig:ratio}, gas with this density and with a temperature of 5000~K
lies close to or within the regime where $t_{\rm cool, fs} < t_{\rm ff}$
(depending on its fractional ionization), and so it is not surprising that 
\citet{BRO01} find that fine-structure cooling is effective and that the gas 
can cool even in the complete absence of $\mHt$.  It is also important to
stress that \citet{BRO01} do not find fine-structure cooling to be important
at densities $n \ll 100 \: {\rm cm^{-3}}$, as the almost adiabatic initial evolution
of the gas makes clear.

For comparison, our simulations start with a high initial gas temperature
of $10^{4} \: {\rm K}$ and involve halos with virial temperatures 
$T_{\rm vir} < 10^{4} \: {\rm K}$; for instance,  $T_{\rm vir} \simeq 1900 \: {\rm K}$
in our fiducial simulation. This means that in our simulations, 
$T_{\rm gas} > T_{\rm vir}$ initially, with the result that thermal pressure support 
is important right from the start. Indeed, at the beginning of our simulations pressure 
precisely balances gravity, by design. Therefore, there is no initial phase of pure 
free-fall collapse as in the \citet{BRO01} simulations. Instead, significant
gravitational contraction occurs only if the gas is able to cool to a temperature of order
$T_{\rm vir}$ or below, which, since fine-structure cooling is initially ineffective, 
will only occur if enough $\mHt$ can form in the low-density gas. At late times 
in our simulations, we would expect to see fine-structure cooling become more 
effective at lower $Z$, in line with what \citet{BRO01} find, but unfortunately the
required gas density is only marginally resolvable with our current simulations.

While the two sets of simulations therefore appear to be consistent with each
other, a key question is obviously which set of initial conditions is more 
appropriate for describing the early evolution of small, metal enriched 
protogalaxies. We argue that it is difficult to see how intergalactic gas could 
become metal enriched without at some point being ionized. Previous 
calculations have shown that the size of a typical
region enriched by a Population III supernova is much smaller than the size of the 
$\hii$ region created by its progenitor star \citep[see e.g.][]{BRO03}. We would therefore 
expect our initial conditions to be more appropriate than those of \citet{BRO01} for
treating recently enriched and ionized regions. However, if enough time elapses 
following the enrichment event for the gas to be able to cool down to a temperature
of a few hundred K, then the Bromm et~al. (2001) initial conditions will be more appropriate.
As \citet{OH03} show, this is most likely to occur in high-redshift gas with a low 
overdensity. Then Compton cooling is fast and highly effective and can cool the 
gas to $T \sim 300 \: {\rm K}$ within a recombination time. On the other hand, at 
lower redshifts, or at higher overdensities, the gas recombines faster than it cools,
and the temperature that can be reached by Compton cooling alone is much higher.
This is the case in our simulated halos. Note that in either case the metals play no 
significant role in determining whether or not the gas is able to collapse.

An important implication of these results is that if we are primarily concerned with 
investigating questions such as how $M_{\mathrm{crit}}$ evolves with redshift,
or how UV feedback in the form of Lyman-Werner photons affects the ability of 
the gas to cool, then we need not worry about the effects of metal enrichment.
This is because the thermal evolution of the gas on the scales of interest for 
these questions is completely dominated by Compton cooling and/or $\mHt$ 
cooling.  Therefore, results from studies such as \citet{HAI00} or \citet{YOS03} 
give a better guide to the behaviour of small, low-metallicity protogalaxies than 
might have been anticipated (although the additional complications posed by 
the mechanical energy injected into the gas by \hii regions and supernovae 
do of course still need to be taken into account). 
 
Finally, it is important to stress that our results do not address the question of 
whether or not there is a critical metallicity ${Z}_{\rm crit}$ above which fine 
structure cooling from metals allows efficient fragmentation to occur. This is 
because if fragmentation does occur, we would expect it to occur at densities
$n > 500 \: {\rm cm^{-3}}$, which are unresolved in our current simulations. We
will examine this question with much higher resolution simulations in a
future paper.

\acknowledgments

We thank Z. Haiman and S. Kitsionas for useful discussions. We also thank
the anonymous referee for his report, which helped us to greatly improve
the presentation of this paper. R.~S.~K. and A.~K.~J. acknowledge support from
the Emmy Noether Program of the Deutsche Forschungsgemeinschaft 
(grant KL1358/1). M.-M.~M.~L. and S.~C.~O.~G. acknowledge support from 
NSF grants AST99-85392 and AST03-07793 and NASA grants NAG5-10103 and 
NAG5-13028. The simulations discussed in this paper were performed on the 
PC cluster ``sanssouci'' at the Astrophysikalisches Institut Potsdam.

\clearpage

\begin{deluxetable}{lcl}
\tablewidth{0pt}
\tablecaption{List of Reactions Included in Our Chemical Network \label{tab:chem_gas}}
\tablehead{Number\  & Reaction & Reference\ } 
\startdata
& & \\
1 & $\mH  + \me  \rightarrow \Hm + \gamma$ & \citet{WIS79} \\
& & \\
2 & $\Hm  + \mH  \rightarrow \mHt + \me$ & \citet{LAU91} \\
& & \\
3 & $\mH  + \Hp  \rightarrow \mHtp + \gamma$ & \citet{RAM76} \\
& & \\
4 & $\mH + \mHtp \rightarrow \mHt + \Hp$ & \citet{KAR79} \\
& & \\
5 & $\Hm  + \Hp  \rightarrow \mH + \mH$ & \citet{MOS70} \\
& & \\
6 & $\mHtp + \me \rightarrow \mH + \mH$ & \citet{SCH94} \\
& & \\
7 & $\mHt + \Hp  \rightarrow \mHtp + \mH$ & \citet{SAV04} \\
& & \\
8 & $\mHt + \me  \rightarrow  \mH + \mH +  \me$ & \citet{STI99} \\
& & \\
9 & $\mHt + \mH  \rightarrow  \mH + \mH + \mH$ & \citet{MAC86} \\
& & \\
10 & $\mHt + \mHt \rightarrow  \mHt + \mH + \mH$ & \citet{MAR98} \\
& & \\
11 & $\mH  + \me  \rightarrow \Hp + \me + \me$ & \citet{JAN87} \\
& & \\
13 & $\Hp  + \me  \rightarrow \mH +  \gamma$ & \citet{FER92} \\
& & \\
15 & $\Hm  + \me  \rightarrow \mH + \me + \me$ &  \citet{JAN87}  \\
& & \\
16 & $\Hm  + \mH  \rightarrow  \mH + \mH +  \me$ &  \citet{JAN87}  \\
& & \\
17 & $\Hm + \Hp   \rightarrow \mHtp + \me$ & \citet{POU78} \\
& & \\
30 & $\cp + \me \rightarrow \mC  + \gamma$ & \citet{NAH97} \\
& & \\
31 & $\sip + \me \rightarrow \mSi + \gamma$ &  \citet{NAH00} \\
& & \\
32 & $\op + \me  \rightarrow \mO + \gamma$ & \citet{NAH99} \\
& & \\
33 & $\mC  + \me  \rightarrow \cp  + \me + \me$ &  \citet{VOR97} \\
& & \\
34 & $\mSi + \me  \rightarrow \sip + \me + \me$ &  \citet{VOR97} \\
& & \\
35 & $\mO  + \me  \rightarrow \op  + \me + \me$ &  \citet{VOR97} \\
& & \\
36 & $\op  + \mH  \rightarrow \mO  + \Hp$ & \citet{STA99} \\
& & \\
37 & $\mO  + \Hp  \rightarrow \op  + \mH$ &  \citet{STA99} \\
& & \\
39 & $\mC  + \Hp  \rightarrow \cp  + \mH$ &  \citet{STA98} \\
& & \\
42 & $\mSi + \Hp  \rightarrow \sip + \mH$ & \citet{KIN96} \\
& & \\
44 & $\cp  + \mSi \rightarrow \mC + \sip$ &  \citet{TEU00} \\
& & \\
51 & $\Hm + \gamma \rightarrow \mH + \me$ & \citet{WIS79} \\
& & \\
52 & $\mHtp + \gamma \rightarrow \mH + \Hp$ & \citet{DUN68} \\
& & \\
53 & $\mHt + \gamma \rightarrow \mH + \mH$ & \citet{DRA96} \\
& & \\
56 & $\mC + \gamma \rightarrow  \cp  + \me$ & \citet{VER96} \\
& & \\
58 & $\mSi + \gamma \rightarrow \sip + \me$ & \citet{VER96} \\
& & \\
60 & $\mH + \mH(+ {\rm grain}) \rightarrow \mHt$ & \citet{HOL79} \\
& & \\
61 & $\Hp + \me(+ {\rm grain}) \rightarrow \mH$  & \citet{WEI01} \\
& & \\
64 & $\cp + \me(+ {\rm grain}) \rightarrow \mC$ & \citet{WEI01} \\
& & \\
66 & $\sip + \me(+ {\rm grain}) \rightarrow \mSi$ & \citet{WEI01} \\
& & \\
\enddata
\tablecomments{The reaction numbering scheme used here is the same 
as that in paper I. References are to the primary source of data for each 
reaction.}
\end{deluxetable}

\begin{deluxetable}{lll}
\tablewidth{0pt}
\tablecaption{List of Photochemical Reaction Rates. \label{tab:chem_gas_photo}}
\tablehead{Number\ & Reaction & Rate $(J_{21}^{-1} \: {\rm s}^{-1})$}
\startdata
51 & $\Hm + \gamma \rightarrow \mH + \me$  & $R_{51} = 1.36 \times 10^{-11}$ \\
52 & $\mHtp + \gamma \rightarrow \mH + \Hp$ & $R_{52} = 4.11 \times 10^{-12}$ \\
53 & $\mHt + \gamma \rightarrow \mH + \mH$ & $R_{53} = 1.30 \times 10^{-12}$ \\
56 & $\mC + \gamma \rightarrow  \cp  + \me$ & $R_{56} = 5.56 \times 10^{-12}$ \\
58 & $\mSi + \gamma \rightarrow \sip + \me$ & $R_{58} = 2.44 \times 10^{-11}$ \\
\enddata
\tablecomments{The reaction numbering scheme used here is the same as that in 
Paper I. Rates are calculated assuming an incident spectrum corresponding to a modified, 
diluted $10^{5} \: {\rm K}$ black body, as described in the text. The value $J_{21}$ quantifies 
the strength of the radiation field at the Lyman limit:  $J(\nu_{\alpha}) = 
10^{-21} J_{21} \: {\rm erg} \: {\rm s^{-1}} \: {\rm cm^{-2}} \: {\rm Hz^{-1}} \: {\rm sr^{-1}}$. 
The rate listed for $\mHt$ photodissociation (reaction 53) is for unshielded gas; our
treatment of $\mHt$ self-shielding is described in section~\ref{chemcool}.}
\end{deluxetable}

\begin{deluxetable}{ll}
\tablecaption{Processes Included in Our Thermal Model. \label{cool_model}}
\tablewidth{0pt}
\tablehead{
\colhead{Process}  & \colhead{References} }
\startdata
{\bf Cooling} & \\
Fine-structure lines ($\mC, \cp, \mO, \mSi, \sip$) & See Paper I \\
Ly$\alpha$ cooling & \citet{CEN92} \\
$\mH$ collisional ionization & \citet{JAN87} \\
Compton cooling & \citet{CEN92} \\
$\Hp$ recombination & \citet{FER92} \\
$\mHt$ rovibrational lines & \citet{BOU99} \\
$\mHt$ collisional dissociation & \citet{MAC86,MAR98} \\
Gas-grain energy transfer$^a$ & \citet{HOL89} \\
{\bf Heating} & \\
Photoelectric effect & \citet{BAK94,WOL95} \\
$\mHt$ photodissociation & \citet{BLA77} \\ 
UV pumping of $\mHt$ & \citet{BUR90}  \\
$\mHt$ formation on dust grains & \citet{HOL89} \\
\enddata
\tablecomments{a: If $T_{\rm gas} < T_{\rm grain}$, the net flow of 
energy is from the grains to the gas, leading to heating instead of
cooling.}
\end{deluxetable}

\begin{deluxetable}{lccccc}
\tablewidth{0pt}
\tablecaption{List of the Runs Discussed in this Paper \label{tab:run_params}}
\tablehead{\colhead{Run} & \colhead{Mass($M_{\odot}$)} & \colhead{Redshift} &
\colhead{$Z / Z_{\odot} $} & \colhead{Dust\tablenotemark{a}} &
\colhead{$J_{21}$\tablenotemark{b}}}
\startdata
S30-Z1 & $4.0 \times 10^{4}$ & 30 & 0.0 & No & 0.0  \\
I30-Z1 & $4.8 \times 10^{5}$ & 30 & 0.0 & No & 0.0 \\
I30-L1 & $4.8 \times 10^{5}$ & 30 & $10^{-3}$ & No & 0.0 \\
I25-Z1-ROT\tablenotemark{c} & $7.8 \times 10^{5}$ & 25 & 0.0 & No & 0.0 \\
I25-Z1 & $7.8 \times 10^{5}$ & 25 & 0.0 & No & 0.0 \\
I25-L1 & $7.8 \times 10^{5}$ & 25 & $10^{-3}$ & No & 0.0 \\
I25-L1D & $7.8 \times 10^{5}$ & 25 & $10^{-3}$ & Yes & 0.0 \\
I25-M1 & $7.8 \times 10^{5}$ & 25 & $10^{-2}$ & No & 0.0 \\
I25-M1D & $7.8 \times 10^{5}$ & 25 & $10^{-2}$ & Yes & 0.0 \\
I25-N1 & $7.8 \times 10^{5}$ & 25 & $10^{-1}$ & No & 0.0 \\
I25-S1 & $7.8 \times 10^{5}$ & 25 & 1.0 & No & 0.0 \\
I25-S1D & $7.8 \times 10^{5}$ & 25 & 1.0 & Yes & 0.0 \\
I25-Z2 & $7.8 \times 10^{5}$ & 25 & 0.0 & No & $10^{-2}$ \\
I25-Z3 & $7.8 \times 10^{5}$ & 25 & 0.0 & No & $10^{-1}$ \\
I25-L3 & $7.8 \times 10^{5}$ & 25 & $10^{-3}$ & No & $10^{-1}$ \\
I25-M3 & $7.8 \times 10^{5}$ & 25 & $10^{-2}$ & No & $10^{-1}$ \\
I25-M3D & $7.8 \times 10^{5}$ & 25 & $10^{-2}$ & Yes & $10^{-1}$ \\
I25-N3 & $7.8 \times 10^{5}$ & 25 & $10^{-1}$ & No & $10^{-1}$ \\
I25-S3 & $7.8 \times 10^{5}$ & 25 & 1.0 & No & $10^{-1}$ \\
I20-Z1 & $1.5 \times 10^{6}$ & 20 & 0.0 & No & 0.0 \\
I20-L1 & $1.5 \times 10^{6}$ & 20 & $10^{-3}$ & No & 0.0 \\
L30-Z1 & $4.8 \times 10^{6}$ & 30 & 0.0 & No & 0.0 \\
L30-L1 & $4.8 \times 10^{6}$ & 30 & $10^{-3}$ & No & 0.0 \\
L20-Z1 & $1.2 \times 10^{7}$ & 20 & 0.0 & No & 0.0 \\
L20-L1 & $1.2 \times 10^{7}$ & 20 & $10^{-3}$ & No & 0.0 \\
\enddata
\tablenotetext{a}{Indicates whether the run includes dust or only gas-phase metals}
\tablenotetext{b}{Strength of the UV background}
\tablenotetext{c}{This run included initial rotation, with a spin
parameter $\lambda = 0.05$}
\end{deluxetable}

\begin{deluxetable}{cccccc}
\tablewidth{0pt}
\tablecaption{Halo Mass Resolution and Virial
   Temperature \label{tab:run_derived_vals}}
\tablehead{\colhead{Halo mass  ($M_{\odot}$)} & \colhead{Redshift} & 
\colhead{Particle number $N$} & \colhead{$M_{\rm part}$($M_{\odot}$)} &
\colhead{$M_{\rm res}$($M_{\odot}$)} & \colhead{$T_{\rm vir}$(K)}}
\startdata
 $4.0 \times 10^{4}$ & 30 & 132651 & 0.25 & 20 & 310 \\
 $4.8 \times 10^{5}$ & 30 & 117649 & 2.0 & 160 & 1600 \\
 $7.8 \times 10^{5}$ & 25 & 140608 & 2.5 & 200 & 1900 \\
 $1.5 \times 10^{6}$ & 20 & 140608 & 5.0 & 400 & 2400 \\
 $4.8 \times 10^{6}$ & 30 & 493039 & 5.0 & 400 & 7600 \\
 $1.2 \times 10^{7}$ & 20 & 1042189 & 5.0 & 400 & 9500 \\
\enddata
\end{deluxetable}

\begin{deluxetable}{lccccccc}
\tablewidth{0pt}
\tablecaption{Physical State of the Densest Gas within the Scale Radius $r_{s}$ 
at Time $t_{\rm end}$ \label{tab:run_results}}
\tablehead{\colhead{Run} & \colhead{$Z$\tablenotemark{a}} & \colhead{$J_{21}$\tablenotemark{b}} &
\colhead{$t_{\rm end}$\tablenotemark{c}} &\colhead{$T_{\rm c, min}$\tablenotemark{d}} & 
\colhead{$n_{\rm c, max}$\tablenotemark{e}} & 
\colhead{$x_{\rm \mHt, c, max}$\tablenotemark{f}} & 
\colhead{$f_{\rm sink}$\tablenotemark{g}}\\\colhead{} & \colhead{($Z_{\odot}$)} & 
\colhead{} &  \colhead{(Myr)} & \colhead{(K)} &
\colhead{(${\rm cm^{-3}}$)} & \colhead{} & \colhead{}}
\startdata
S30-Z1 & $0.0$ & $0.0$ & 100 & 1100 & 0.3 & $9 \times 10^{-4}$ & 0.0\\
I30-Z1   & $0.0$ & $0.0$ & 75 & $< 200 $ & $> 500$ & $5 \times 10^{-3}$ & 0.04\\
I30-L1   & $10^{-3}$ & $0.0$ & 100 & $< 200 $ & $> 500$ & $3 \times 10^{-3}$ & 0.19\\
I25-Z1-ROT & $0.0 $ & $0.0 $ & 100 & $< 200 $ & $> 500$ & $4 \times 10^{-3}$ & $2 \times 10^{-3}$\\
I25-Z1 & $0.0 $ & $0.0 $ & 100 & $< 200 $ & $> 500$ & $4 \times 10^{-3}$ & 0.17\\
I25-L1 & $10^{-3} $ & $0.0 $ & 100 & $< 200 $ & $> 500$ & $2 \times 10^{-3}$ & 0.20\\
I25-L1D & $10^{-3} $ & $0.0 $ & 100 & $< 200 $ & $> 500$ & $0.5$ & 0.13\\
I25-M1 & $10^{-2} $ & $0.0 $ & 100 & $< 200 $ & $> 500$ & $3 \times 10^{-3}$ & 0.15\\
I25-M1D & $10^{-2} $ & $0.0 $ & 100 & $< 200 $ & $> 500$ & $0.5$ & 0.08\\
I25-N1 & $10^{-1} $ & $0.0 $ & 100 & $< 200 $ & $> 500$ & $2 \times 10^{-3}$ & 0.17\\
I25-S1 & $1.0$ & $0.0 $ & 100 & $< 200 $ & $> 500$ & $1 \times 10^{-3}$ & 0.28\\
I25-S1D & $1.0$ & $0.0 $ & 100 & $< 200 $ & $> 500$ & $0.5$ & 0.27\\
I25-Z2 & $0.0 $ & $10^{-2} $ & 100 & $< 200 $ & $> 500$ & $5 \times 10^{-4}$ & $5 \times 10^{-3}$\\
I25-Z3 & $0.0 $ & $10^{-1} $ & 100 & $4400 $ & $0.7 $ & $1.6 \times 10^{-5}$ & 0.0\\
I25-L3 & $10^{-3} $ & $10^{-1} $ & 100 & $5900 $ & $0.6$ & $1.5 \times 10^{-5}$ & 0.0\\
I25-M3 & $10^{-2} $ & $10^{-1} $ & 100 & $5600$ & $0.8$ & $1.7 \times 10^{-5}$ & 0.0\\
I25-M3D & $10^{-2}$ & $10^{-1}$ & 100 & 5900 & 0.7 & $1.5 \times 10^{-5}$ & 0.0\\
I25-N3 & $10^{-1} $ & $10^{-1} $ & 100 & $2600 $ & $2.5$ & $5 \times 10^{-5}$ & 0.0\\
I25-S3 & $1.0$ & $10^{-1} $ & 100 & $< 200 $ & $> 500$ & $5 \times 10^{-4}$ & 0.3\\
I20-Z1   & $0.0$ & $0.0$ & 100 & 160 & 1600 & $3 \times 10^{-3}$ & 0.0\\
I20-L1   & $10^{-3}$ & $0.0$ & 100 & 130 & 2100 & $3 \times 10^{-3}$ & 0.0\\
L30-Z1   & $0.0$ & $0.0$ & 25 & $< 200 $ & $> 500$ & $7 \times 10^{-3}$ & $1 \times 10^{-3}$\\ 
L30-L1   & $10^{-3}$ & $0.0$ & 10 & $< 200 $ & $> 500$ & $3 \times 10^{-3}$ & $2 \times 10^{-4}$\\
L20-Z1   & $0.0$ & $0.0$ & 20 & $< 200 $ & $> 500$ & $7 \times 10^{-3}$ & $6 \times 10^{-4}$\\
L20-L1   & $10^{-3}$ & $0.0$ & 10 & 310 & 330 & $3 \times 10^{-3}$ & 0.0 \\
\enddata
\tablenotetext{a}{Metallicity of the gas.}
\tablenotetext{b}{Strength of the UV background.}
\tablenotetext{c}{Time at the end of the simulation, or $t = 100 \: {\rm Myr}$, whichever
is smaller}
\tablenotetext{d}{Minimum temperature of the gas within the scale radius~$r_{s}$.}
\tablenotetext{e}{Maximum number density of the gas within the scale radius~$r_{s}$.}
\tablenotetext{f}{Maximum fractional H$_2$ abundance within the scale radius~$r_{s}$.}
\tablenotetext{g}{Mass fraction of gas within a sink particle.} 
\end{deluxetable}

\clearpage

\end{document}